\documentclass[structabstract]{aa}  
\usepackage{graphicx}
\usepackage{txfonts}
\usepackage[normalem]{ulem}

\begin{document}

\title{Multiple water band detections in the CARMENES near-infrared transmission spectrum of HD\,189733\,b}
%
%
\titlerunning{Multiple H$_2$O band detections in HD\,189733\,b with CARMENES}
\author{F.~J. Alonso-Floriano\inst{1}, A. S\'anchez-L\'opez\inst{2}, I.~A.~G. Snellen\inst{1}, M. L\'opez-Puertas\inst{2}, E. Nagel\inst{3}, P.~J. Amado\inst{2},  F.~F.~ Bauer\inst{2}, J.~A. Caballero\inst{4}, S. Czesla\inst{3}, L. Nortmann\inst{5,6}, E. Pall{\'e}\inst{5,6}, M. Salz\inst{3}, A. Reiners\inst{7}, I. Ribas\inst{8,9},  A.~Quirrenbach\inst{10}, J. Aceituno\inst{11}, G. Anglada-Escud{\'e}\inst{12}, V.~J.~S.~B{\'e}jar\inst{5,6}, E.~W. Guenther\inst{13}, T. Henning\inst{14},   A.~Kaminski\inst{10}, M. K{\"u}rster\inst{14}, M.~Lamp{\'o}n\inst{2}, L. M. Lara\inst{2}, D.~Montes\inst{15}, J.~C. Morales\inst{8,9},   L. Tal-Or\inst{16,7}, J. H. M. M. Schmitt\inst{3}, M.~R.~Zapatero~Osorio\inst{4}, and M. Zechmeister\inst{5}}
\institute{Leiden Observatory, Leiden University, Postbus 9513, 2300 RA, Leiden, The Netherlands
\and
Instituto de Astrof{\'i}sica de Andaluc{\'i}a (IAA-CSIC), Glorieta de la Astronom{\'i}a s/n, 18008 Granada, Spain
\and
Hamburger Sternwarte, Universit{\"a}t Hamburg, Gojenbergsweg 112, 21029 Hamburg, Germany
\and
Centro de Astrobiolog{\'i}a, CSIC-INTA, ESAC campus, Camino bajo del castillo s/n, 28692 Villanueva de la Ca{\~n}ada, Madrid, Spain
\and
Instituto de Astrof{\'i}sica de Canarias (IAC), Calle V{\'i}a Lactea s/n, E-38200 La Laguna, Tenerife, Spain
\and
Departamento de Astrof{\'i}sica, Universidad de La Laguna, 38026  La Laguna, Tenerife, Spain
\and
Institut f{\"u}r Astrophysik, Georg-August-Universit{\"a}t, 37077 G{\"o}ttingen, Germany
\and
Institut de Ci\`encies de l'Espai (CSIC-IEEC), Campus UAB, c/ de Can Magrans s/n, 08193 Bellaterra, Barcelona, Spain
\and
Institut d'Estudis Espacials de Catalunya (IEEC), 08034 Barcelona, Spain
\and
Landessternwarte, Zentrum f\"ur Astronomie der Universit\"at Heidelberg, K\"onigstuhl 12, 69117 Heidelberg, Germany
\and
Centro Astron{\'o}nomico Hispano Alem{\'a}n, Observatorio de Calar Alto, Sierra de los Filabres, E-04550 G{\'e}rgal, Spain
\and
School of Physics and Astronomy, Queen Mary, University of London, 327 Mile End Road,
London, E1 4NS, UK
\and
Th{\"u}ringer Landessternwarte Tautenburg, Sternwarte 5, 07778 Tautenburg, Germany
\and
Max-Planck-Institut f{\"u}r Astronomie, K{\"o}nigstuhl 17, 69117 Heidelberg, Germany
\and
Departamento de F{\'i}sica de la Tierra y Astrof{\'i}sica, Facultad de Ciencias F{\'i}sicas, Universidad Complutense de Madrid, 28040 Madrid, Spain
\and
School of Geosciences, Raymond and Beverly Sackler Faculty of Exact Sciences, Tel Aviv University, Tel Aviv, 6997801, Israel
}
\authorrunning{F.~J. Alonso-Floriano et al.}
\date{Received 28 Sep 2018; accepted 03 Nov 2018}

\abstract
{}
{We explore the capabilities of CARMENES for characterizing hot-Jupiter atmospheres by targeting multiple water bands, in particular, those at 1.15 and 1.4\,$\mu$m. 
{\it Hubble Space Telescope} observations suggest that this wavelength region is relevant for distinguishing between hazy/cloudy and clear atmospheres.}
{We observed one transit of the hot Jupiter HD 189733 b with CARMENES. 
Telluric and stellar absorption lines were removed using {\sc Sysrem}, which performs a principal component analysis including proper error propagation. The residual spectra were analysed for water absorption  with cross-correlation techniques using synthetic atmospheric absorption models. }
{We report a cross-correlation peak at a signal-to-noise ratio (SNR) of 6.6, revealing the presence of water in the transmission spectrum of HD\,189733\,b. The absorption signal appeared slightly blueshifted at --3.9\,$\pm$\,1.3\,km\,s$^{-1}$. We measured the individual cross-correlation signals of the water bands at 1.15 and 1.4\,$\mu$m, finding cross-correlation peaks at SNRs of 4.9 and 4.4, respectively. The 1.4\,$\mu$m feature is consistent with that observed with the \textit{Hubble Space Telescope}.  }
{The water bands studied in this work have been mainly observed in a handful of planets from space. The ability of also detecting them individually from the ground at higher spectral resolution can provide insightful information to constrain the properties of exoplanet atmospheres.
Although the current multiband detections can not yet constrain atmospheric haze models for HD\,189733\,b, future observations at higher signal-to-noise ratio could provide an alternative way to achieve this aim.}
\maketitle
%

\section{Introduction}
\label{section.introduction}

Shortly after the discovery of the first hot Jupiter \citep{Mayor95}, the theoretical basis for detecting exoplanet atmospheres through transmission spectroscopy was developed \citep{Seager00,Brown01,Hubbard01}. 
Subsequently, \citet{Charbonneau02} were the first to measure the transmission signal from sodium in the atmosphere of HD 290458\,b using the Space Telescope Imaging Spectrograph on the \textit{Hubble Space Telescope} ($HST$). For a long time it was thought that atmospheric exoplanet science could only be conducted from space, either with the $HST$ or the \textit{Spitzer Space Telescope}, the latter detecting the first thermal emission from exoplanets \citep{Charbonneau05, Deming05b}. It was not until the publications by \citet{Redfield08} and \citet{Snellen08} that ground-based observations started to play a role. Both studies presented detections of sodium in HD\,189733\,b and HD\,209458\,b, respectively, using high-dispersion spectrographs at 8-10-m class telescopes. 

In earlier works (e.g. \citealt{Charbonneau99,Collier-Cameron99,Deming05a,Rodler08,Rodler10}) it was already envisaged that ground-based high-dispersion spectroscopy could be a powerful tool to characterise exoplanet atmospheres, either in reflected light, intrinsic thermal emission, or in transmission. 
In the case of reflection, the light scattered from the planetary atmosphere contains a faint copy of the stellar spectrum, but Doppler-shifted depending on the changing radial component of the orbital velocity of the planet, which can be as large as 150\,km\,s$^{-1}$ or more for close-in planets. 
The stationary stellar component can be filtered out, after which the residual spectra contain the Doppler-shifted component reflected off the planet. In the same way molecular lines can be probed in the planetary thermal or transmission spectrum. 
Due to the planet/star contrast the technique is more efficient at infrared wavelengths, where it is used to remove the telluric absorption in a similar way as the stellar lines in the optical.
A first detection was presented by \cite{Snellen10}, who measured carbon monoxide in the transmission spectrum of HD 209458\,b using the Cryogenic InfraRed {\'E}chelle Spectrograph (CRIRES; \citealt{Kaeufl04}) at the ESO Very Large Telescope (VLT). 
By measuring the radial velocity of the planet, the system could be treated as an edge-on spectroscopic binary system allowing for an assumption-free mass determination of both the star and the planet. Also, a marginally significant blue shift of 2\,$\pm$\,1\,km\,s$^{-1}$ of the CO signal was interpreted as coming from the radial velocity component of a possible global wind blowing from the hot day-side to the cold night-side of the planet. 

\begin{figure*}
\centering
\includegraphics[angle= 0, width=\textwidth,trim={40 40 0 0}]{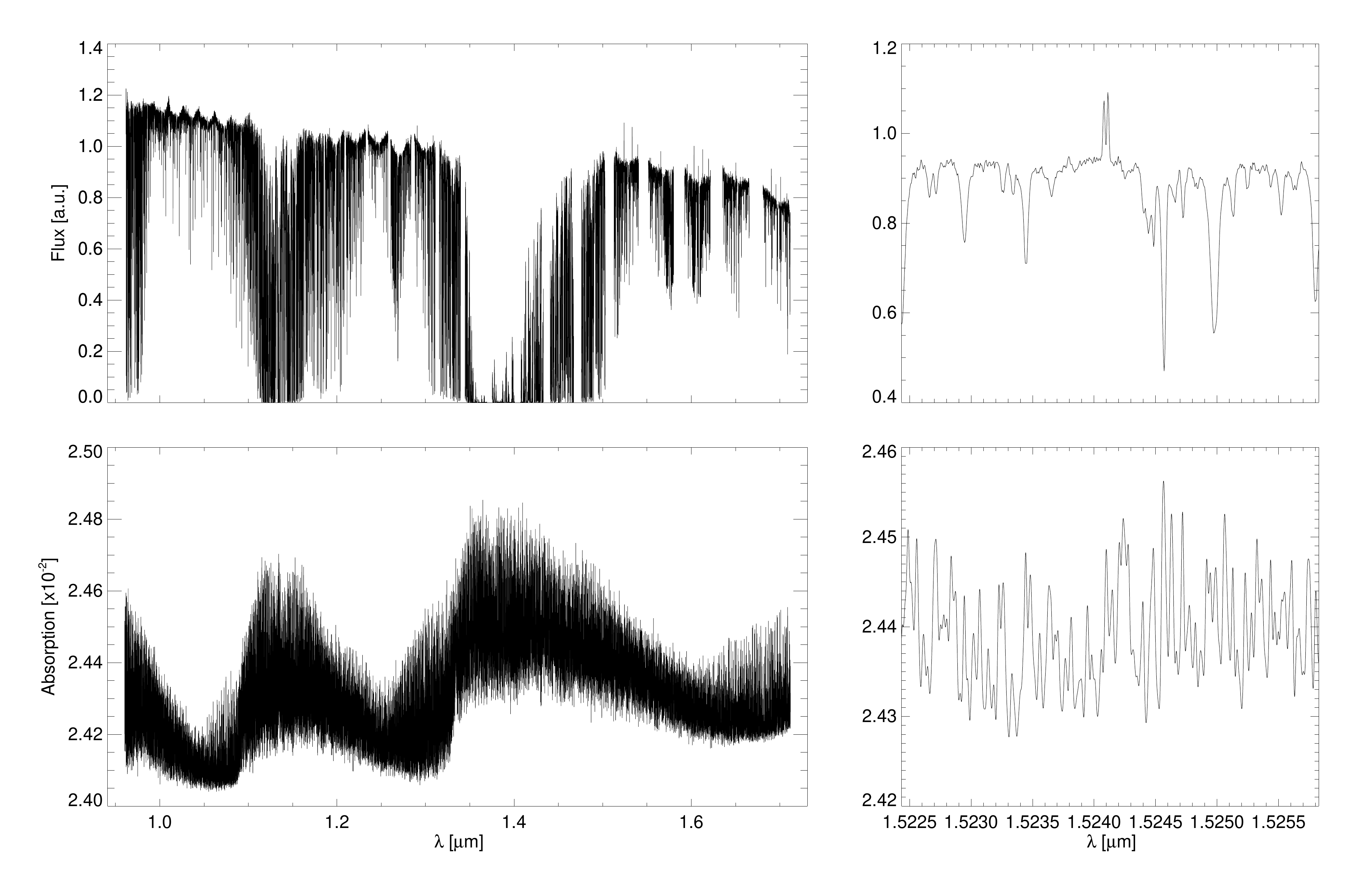}
\caption{\label{fig.spectra_and_model}
{\em Top panels}: CARMENES NIR spectrum of HD\,189733\,b taken on 2017 September 07 at 22:06:29\,UT ({\em left panel}) and a zoom-in on order 40 around 1.524\,$\mu$m ({\em right panel}). 
The major telluric absorptions by water vapour bands near 0.95, 1.14 and 1.4\,$\mu$m are prominent. For this plot, the original spectrum, as provided by the CARMENES pipeline, was approximately corrected for the instrument response using an early type telluric standard star from a previous observation; so the decline of the continuum level is mainly due to the decrease of the stellar flux. The V shapes of the continuum level within each order are instrumental effects, either due to overestimating the blaze function or the background illumination. This effect is mainly caused by the different observational conditions of the calibration star and the target, such as target altitude and background illumination. In our analysis, the instrument response was not applied, instead, we normalized each order of the spectra.
The emission features observed at the reddest wavelengths (>1.5\,$\mu$m) in the left and right panels  are telluric emission lines.
{\em Bottom panels}: Atmospheric absorption model of HD\,189733\,b in the CARMENES NIR spectral range ({\em left panel}) and a zoom-in around 1.524\,$\mu$m ({\em right panel}). The contribution of the planetary disk (i.e., opaque at all wavelengths) has been included. The most prominent features are caused by the atmospheric H$_{2}$O bands. The calculations were performed with the pressure-temperature profile of \citet{Brogi18} (see Fig.~\ref{fig.pt}) and a constant H$_{2}$O volume mixing ratio of 10$^{-4}$. The model was computed at a very high spectral resolution ($\mathcal{R}$\,$\sim$\,4$\cdot$10$^{7}$) and convolved with the CARMENES line spread function. 
}
\end{figure*}

At a resolving power of $\mathcal{R}$\,=\,100\,000, molecular bands are resolved in tens to hundreds of individual lines (see Fig.\,\ref{fig.spectra_and_model}). 
While the observing technique filters out any possible broad-band feature (and, with it, most instrumental calibration errors), the signatures of the individual lines are preserved. Because they are mostly too faint to be detected individually, cross-correlation techniques are used to optimally combine the lines to produce a joint molecular signal.  
The closer the model template matches the planet spectrum, the stronger the resulting cross-correlating signal should be. 
In principle, this can be used to constrain the temperature structure of the planet atmosphere and the volume mixing ratio of the targeted molecule. However, the sensitivity to these parameters is low, as there is a known degeneracy when comparing the models to the data (e.g., \citealt{Brogi14}.)
The technique has since thrived, with detections of both, water and carbon monoxide, in the atmospheres of a handful of planets, in transmission and emission, using CRIRES at the VLT (e.g., \citealt{Brogi12,Rodler12,Birkby13,deKok13,Brogi13,Brogi14,Schwarz15,Schwarz16,Brogi16,Birkby17}) and the Near-Infrared Spectrograph on the Keck\,II Telescope \citep{Lockwood14,Piskorz16,Piskorz17}. 

Ground-based high-dispersion spectroscopy has recently ventured out in several different directions. \cite{Nugroho17} detected, for the first time,  titanium oxide (TiO) in the day-side thermal emission spectrum of WASP-33\,b, probed in the optical wavelength regime using the High Dispersion Spectrograph on the Subaru Telescope. They showed the molecular lines to be in emission, which conclusively proved the presence of a thermal inversion. Also, improved instrumental stability and analysis techniques now allow the use of smaller telescopes. \cite{Wyttenbac15} and \cite{Louden15} used data from HARPS (High Accuracy Radial-velocity Planet Searcher {\'e}chelle spectrograph) at the ESO 3.6 m telescope \citep{Mayor03} to measure in detail the shape  of the sodium transmission signature of HD\,189733\,b.  \cite{Brogi18} used GIANO at the Telescopio Nazionale Galileo to detect the presence of water in the atmosphere of HD\,189733\,b. 
\cite{Yan18} used the highly-stabilised high-dispersion spectrograph CARMENES (Calar Alto high-Resolution search for M dwarfs with Exoearths with Near-infrared and optical {\'E}chelle Spectrographs; \citealt{Quirrenbach16}) to detect atomic hydrogen absorption (H$\alpha$) during the transit of Kelt-9b. Also in this target, \cite{Hoeijmakers18} used HARPS-North transit observations to find for the first time absorption features of atomic heavy elements (Fe, Fe$^{+}$, and Ti$^{+}$).

In this paper, we present transmission spectroscopy of HD\,189733\,b using CARMENES observations targeting its molecular water signature. 
The planet is one of the archetypical hot Jupiters, because it was for long the brightest known transiting system ($V$\,=\,7.6\,mag, $J$\,=\,6.1\,mag).
It has been the target of many observational studies, such as of its hydrogen exosphere \citep{Lecavelier10, Lecavelier12,Bourrier13}, its transmission spectrum (e.g., \citealt{Redfield08,Sing11,Gibson12,McCullough14,Brogi16,Brogi18}), and its emission spectrum \citep{Deming06,Grillmair08,Charbonneau08,Swain10,Todorov14}. 

Our goal is the detection of water in the atmosphere of HD\,189733\,b using the wavelength region covered by the CARMENES NIR channel.
We focus mainly on the two strongest water features near 1.15 and 1.4\,$\mu$m. 
Especially, we attempt the individual detection of these features, to confirm from the ground and with a different technique the 1.4\,$\mu$m water feature as observed with the Wide Field Camera 3 (WFC3) at the $HST$ in several planets (e.g., \citealt{Sing16}).
One of the striking features of the transmission spectrum of HD189733\,b is a blueward slope, most prominent in the ultraviolet and blue optical, attributed to Rayleigh scattering from haze particles high in the planet atmosphere \citep{Lecavelier08,Pont08,Gibson12,Pont13,Sing16,Pino18a}. The transmission spectra of a small sample of hot Jupiters observed with $HST$ suggest that the water features in the most hazy atmospheres, like that of HD189733\,b, are partly masked.
The comparison between the strength of water absorption over different bands could constrain the Rayleigh scattering of the high-altitude hazes in this kind of planets \citep{Stevenson2016}. Although the known degeneracy on the models make it a complex problem (see \citealt{Heng2017}), the recent work of \citet{Pino18b} suggests that ground-based high-resolution observations over multiple water bands could constrain the presence of hazes in the atmosphere of planets like HD\,189733\,b. In addition, the combination of low-dispersion spectroscopy from space and high-dispersion spectroscopy from the ground over multiple bands provides better constraints to the models \citep{Brogi17}.

\begin{table}
\centering
\caption{\label{table.parameters}Parameters of the exoplanet system HD\,189733.}
\begin{tabular}{l c l} 
   \hline
   \hline
   \noalign{\smallskip}
Parameter					 & Value & Reference  \\
  \noalign{\smallskip}
    \hline
      \noalign{\smallskip}
$\alpha$ [J2000]	& 20:00:43.71 	& {\em Gaia} DR2$^{a}$	\\
\noalign{\smallskip}
$\delta$ [J2000]	& +22:42:35.2	& {\em Gaia} DR2$^{a}$ \\
\noalign{\smallskip}
$V$			& 7.65 mag	&	\citet{Koen10} \\
\noalign{\smallskip}
$J$			& 6.07 mag	&	\citet{Skrutskie06} \\
\noalign{\smallskip}
$R_{\star}$$^{b}$		& 0.756\,$\pm$\,0.018\,$R_\sun$& \citet{Torres08}	\\
\noalign{\smallskip}
$K_{\star}$ 		& 201.96\,$^{+1.07}_{-0.63}$\,m\,s$^{-1}$	& \citet{Triaud09}	\\
\noalign{\smallskip}
$\varv_{\rm sys}$ 	& --2.361\,$\pm$\,0.003\,km\,s$^{-1}$	& \citet{Bouchy05}	\\
\noalign{\smallskip}
$a$		& 0.03120\,(27)\,au	& \citet{Triaud09} \\
\noalign{\smallskip}
$e$		& 0.0041$^{+0.0025}_{-0.0020}$	& \citet{Triaud09} \\
\noalign{\smallskip}
${P_{\rm orb}}$ & 2.21857567\,(15)\,d & \citet{Agol10} \\
\noalign{\smallskip}
$T_{\rm 0}$ & 2454279.436714\,(15)\,d & \citet{Agol10} \\
\noalign{\smallskip}
{\em i} & 85.71\,$\pm$\,0.024\,deg & \citet{Agol10} \\
\noalign{\smallskip}
$R_{\rm P}$$^{b}$ 	& 1.138\,$^{+0.027}_{-0.027}$\,$R_{J}$	& \citet{Torres08}	\\
\noalign{\smallskip}
$K_{\rm P}$$^{c}$ 		& 152.5$^{+1.3}_{-1.8}$\,km\,s$^{-1}$ 	& \citet{Brogi16}	\\
\noalign{\smallskip}
$\varv_{\rm wind}$ 	& --3.9\,$\pm$\,1.3\,km\,s$^{-1}$	&	This work\\
\noalign{\smallskip}
\hline
\end{tabular}
\tablefoot{
\tablefoottext{a}{\citet{Gaia18}.}
\tablefoottext{b}{Equatorial radii.}
\tablefoottext{c}{Value derived from orbital parameters}.}
\end{table}

\section{Observations} \label{section.observations}

We used CARMENES \citep{Quirrenbach16} to observe the system HD\,189733 (Table\,\ref{table.parameters}) on 7 September 2017\footnote{The reduced spectra can be downloaded from the Calar Alto archive, http://caha.sdc.cab.inta-csic.es/calto/}, after two earlier attempts for which the instrument presented stability issues. 
The two spectrograph channels of CARMENES are located in the coud{\'e} room of the 3.5\,m telescope at the Calar Alto Observatory, fed by fibres connected to the front-end mounted on the telescope. One of the CARMENES spectrographs,  dubbed VIS channel, covers the optical wavelength range,  $\Delta\lambda$\,=\,520--960\,nm, through 55 orders, and the other, the NIR channel, covers the near-infrared wavelength range, $\Delta\lambda$\,=\,960--1710\,nm, in 28 orders. The resolving power is $\mathcal{R}$\,=\,94\,600 in the VIS channel and $\mathcal{R}$\,=\,80\,400 in the NIR channel. The detector in the optical is a 4096\,$\times$\,4096 pixel e2v\,231-84 CCD, the near-infrared wavelength range is covered by two 2048\,$\times$\,2048 pixel HAWAII-2RG infrared detectors, which results in a small gap due to the central separation between detectors. Overlap between orders is lost from the H band on, producing further gaps that increase from near-continuous to 15\,nm wide ($\sim$30\% of the order wavelength range) at the long wavelength end (see Fig.\,\ref{fig.spectra_and_model}, top left panel). Only the NIR channel data are considered in this work, since no data were taken with the VIS channel due to shutter problems during our observing night. 

CARMENES offers two fibres, fibre A for the target, and fibre B for either the Fabry-P\'erot etalon or the sky \citep{Seifert12,Sturmer14}. 
The Fabry-P\'erot etalon is used for simultaneous wavelength calibration, in particular for high-precision radial velocity studies. An atlas of calibrated CARMENES spectra can be found in \citet{Reiners18}. However, since the wavelength-calibration requirement for transmission spectroscopy is relatively low, we opted to use the second fibre for sky measurements. These sky measurements were not used in the data calibration process, but only to visually identify sky emission lines. 
The observations were obtained in service mode, consisting of 46 exposures of 198 seconds, starting at 20:14 UT and ending at 00:00 UT, corresponding to a planet orbital phase range of $\phi$\,=\,--0.035 to +0.036.  
During the course of the observation, the relative humidity varied between 57\% and 65\%. 
The star moved from airmass 1.08 to 1.30, during which we updated the atmospheric dispersion corrector four times by reacquisition of the target (which takes about two minutes, (c.f., \citealt{Seifert12}). 
A typical continuum SNR of $\gtrsim$150 was reached per spectrum.

\section{Data reduction} \label{section.reduction}
The CARMENES data were automatically processed after the observations using the dedicated data reduction pipeline 
 {\sc Caracal} v2.10 \citep{Zechmeister2014,Caballero2016}. 
The precipitable water vapour (PWV) was retrieved from the measured spectra using the ESO MOLECFIT tool \citep{Smette2015}.
The PWV level during the observations was relatively high, ranging from  11.7 to 15.9\,mm, compared to an average of $\sim$7\,mm at Calar Alto. This made the telluric water absorption in orders 45--42 (1.34--1.47\,$\mu$m) and 54--53 (1.12--1.16\,$\mu$m) too high for useful data analysis.  

\subsection{Outlier rejection and normalization \label{subsection.aligment}}

The spectra were processed and analysed independently per order, using  very similar techniques to those developed in previous works (\citealt{Birkby17,Brogi16,Brogi17,Brogi18}). First, we removed from the data artefacts attributable to cosmic rays (5\,$\sigma$ outliers) and a small number of pixels flagged as bad quality by the pipeline. 
The pipeline is optimised to obtain long-term radial velocity stability utilising the simultaneously obtained Fabry-P\'erot data. 
Although we did not use this mode, we computed a night drift of $\sim$15\,m\,s$^{-1}$ ($\sim$0.011\,pixel) by measuring the radial velocity variation of telluric lines present in the observed spectra. 
This drift was so small that a wavelength correction was not necessary.
Subsequently, we normalised the spectra using a quadratic polynomial fit to the pseudo-continuum avoiding the sky emission lines, which were present mostly in the reddest orders.  

\subsection{Stellar and telluric signal subtraction using \sc{Sysrem} \label{subsection.telluric}}

\begin{figure*}
\includegraphics[width=\textwidth]{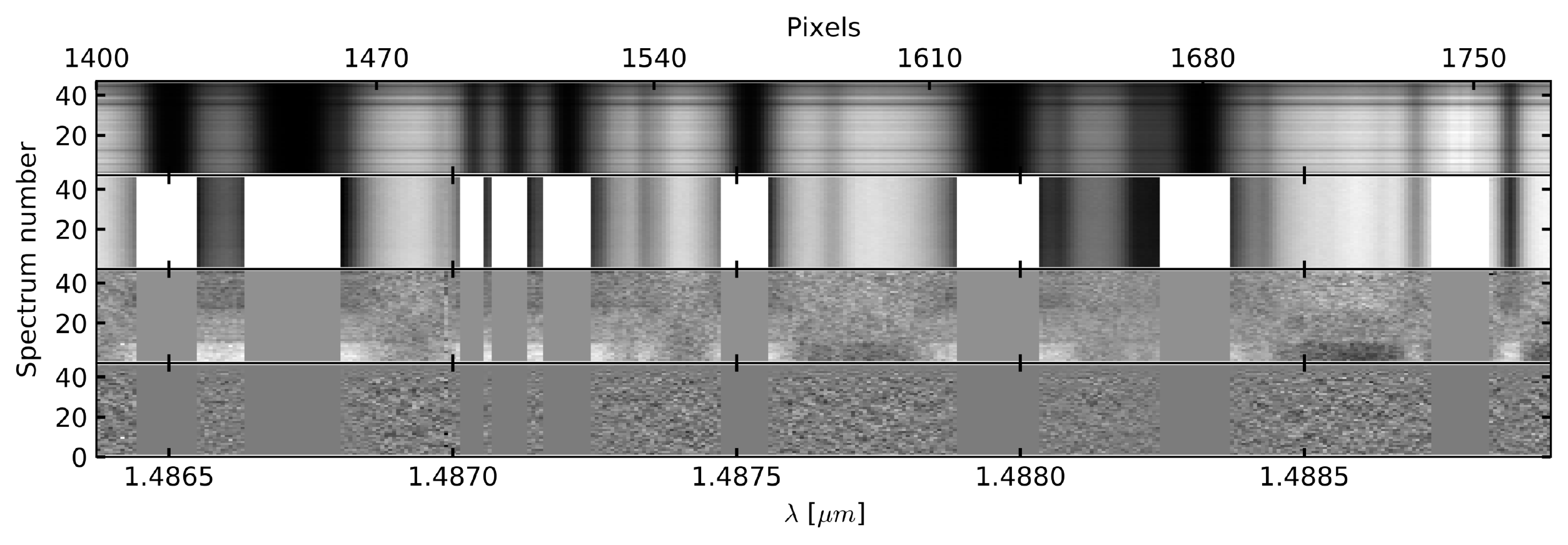}
\caption{\label{fig.matrices_sysrem}. Observed spectra matrices at different steps of the analysis for a short wavelength range in order 41. We plot wavelength along the horizontal axis and the time series of spectra along the vertical axes. {\it Top panel}: matrix of spectra as provided by the CARMENES pipeline. {\it Upper-middle panel}: normalized and masked spectra ready for the telluric and stellar removal process with {\sc Sysrem}. {\it Bottom-middle panel}: matrix of residuals after the first iteration. At this point the telluric residuals are still visible. {\it Bottom panel}: matrix of residuals after the tenth iteration. The telluric residuals are almost completely removed.} 
\end{figure*}

The expected water signature from the planet is several orders of magnitude smaller than the stellar and telluric absorption lines. 
Hence, these contaminations have to be removed before the planetary signal can be detected. Since we do not expect to retrieve any planet signal from the cores of the strongest telluric absorption lines, we masked all wavelengths that exhibit an absorption greater than 80\% of the flux continuum.
We also masked the wavelengths corresponding to strong sky emission lines (see Fig.\,\ref{fig.matrices_sysrem}).  
In total, we masked $\sim$10\% of the spectra.
The remaining stellar and telluric components were removed using {\sc Sysrem} (\citealt{Tamuz05, Mazeh07}), an algorithm that deploys an iterative principal component analysis allowing for unequal uncertainties for each data point. 
This code was already successfully applied for the detection of planet atmospheres by \cite{Birkby17}, \cite{Nugroho17}, and \cite{Hawker2018}. 

We fed {\sc Sysrem} with 22 matrices containing the data of each usable order and ran it, on each of these matrices individually, for 15 iterations. 
We show a zoom-in of one of these matrices in the upper-middle panel of Fig.\,\ref{fig.matrices_sysrem}. The rows of the matrices are composed of the normalized and masked spectra.
Each column of the matrices is a wavelength channel (i.e., pixel column), similar to a "light curve".
{\sc Sysrem} searches and subtracts iteratively common modes between the columns of the matrix, e.g., due to airmass variations.
Thus, the code primarily removes the largest spectral variations, which are the quasi-static stellar and telluric signals, as shown after the first {\sc Sysrem} iteration in Fig.\,\ref{fig.matrices_sysrem}.
Since the planet signal is Doppler-shifted, it is preserved buried in the noise of the resulting residual matrices.
However, {\sc Sysrem} is also prone to remove the planet signal after a few iterations following the removal of the higher order variations.
This effect was studied in detail by \citet{Birkby17} and can start at a different iteration in each order.
We found this also to be dependent on the assumed transmission spectrum model, which is further discussed in Sect.\,\ref{subsection.significance} where we define the optimal iteration to halt the code.

\begin{figure}
\hspace*{0.35cm}\includegraphics[angle=0, width=0.45\textwidth]{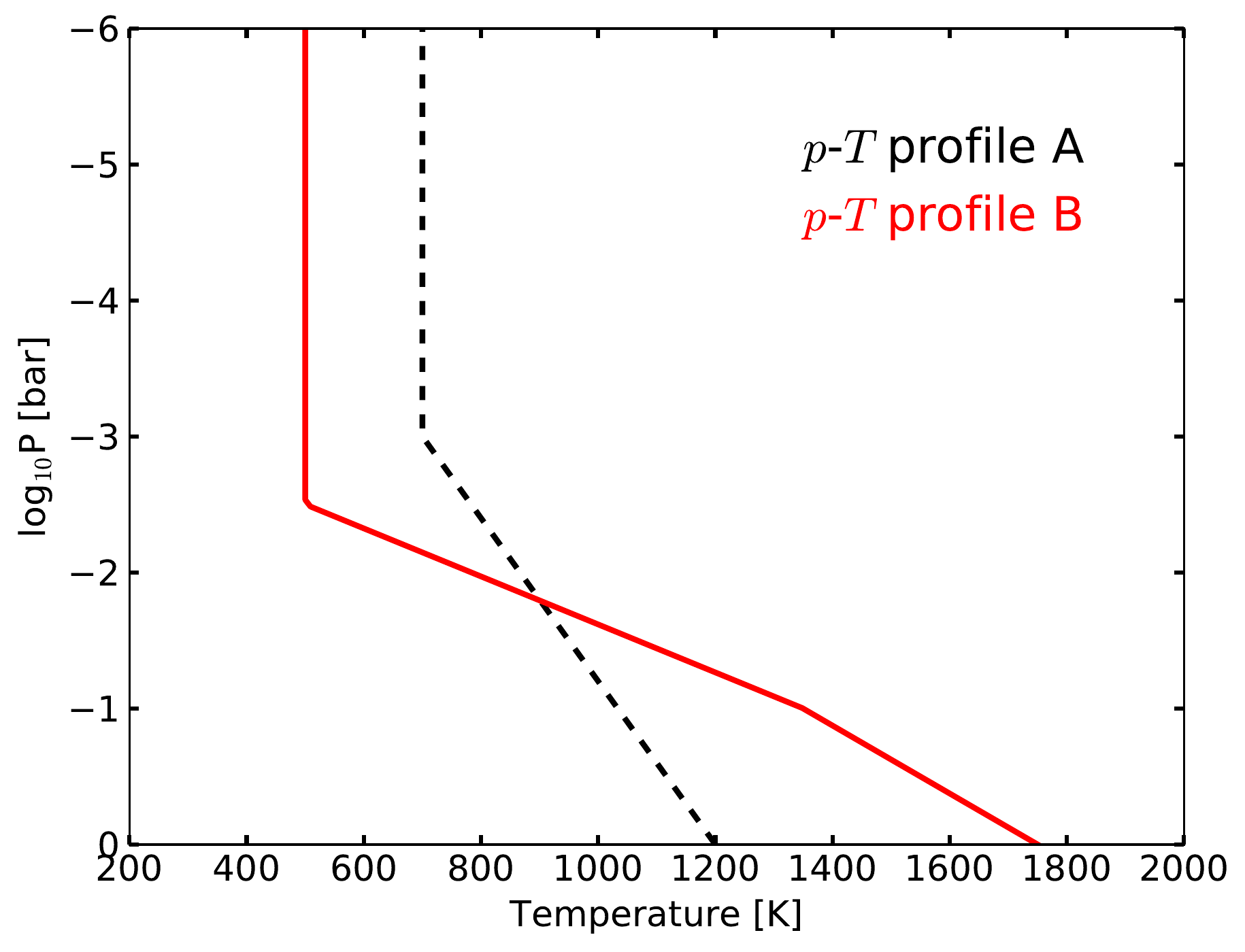}
\caption{\label{fig.pt}Pressure--temperature ($p$--$T$) profiles used for computing the synthetic atmospheric absorption spectra. In dashed black, the $p$--$T$ profile A, was obtained from our best fit to the $HST$/WFC3 data of \cite{McCullough14}. In solid red, the $p$--$T$ profile B was taken from \citet{Brogi18}.} 
\end{figure}

\section{Planet signal retrieval} \label{section.signal_retrieval}

After removing the main telluric and stellar contributions using {\sc Sysrem}, the residual matrices contain the information of thousands of lines belonging to molecular species present in the planet atmosphere, but none of these lines are individually detectable.
However, we can detect the combined signal of these numerous lines by cross-correlating the residual matrices and high-resolution absorption models of the planet atmosphere.

In the wavelength region covered by the CARMENES NIR channel, the main molecular absorption bands expected for HD\,189733\,b are those of water vapour (H$_{2}$O, see Fig.\,\ref{fig.spectra_and_model}). 
Other gases potentially contributing in this spectral region are carbon monoxide (CO) and methane (CH$_4$).
Although the expected abundance of methane is very low \citep{Brogi18}, we tested in our analysis models including only water, only methane, and a combination of water and methane. 
There is a weak CO band approximately from 1.56 to 1.60\,$\mu$m, which is covered by three orders of CARMENES. Based on the results of \cite{Brogi16}, who aimed for the detection of the much stronger CO band from 2.28 to 2.34\,$\mu$m with CRIRES, the detection of CO alone would not be expected in our data. In addition, probing the CO band would require a dedicated correction for the stellar CO residuals left by the Rossiter-McLaughlin effect \citep{Brogi16,Brogi18}. Therefore, we did not include CO in our work, but suggest it for a dedicated study. 
The grid of used absorption models is described in the next section.

\begin{figure*}
\centering
\includegraphics[angle= 0, width=0.45\textwidth, height=0.22\textheight]{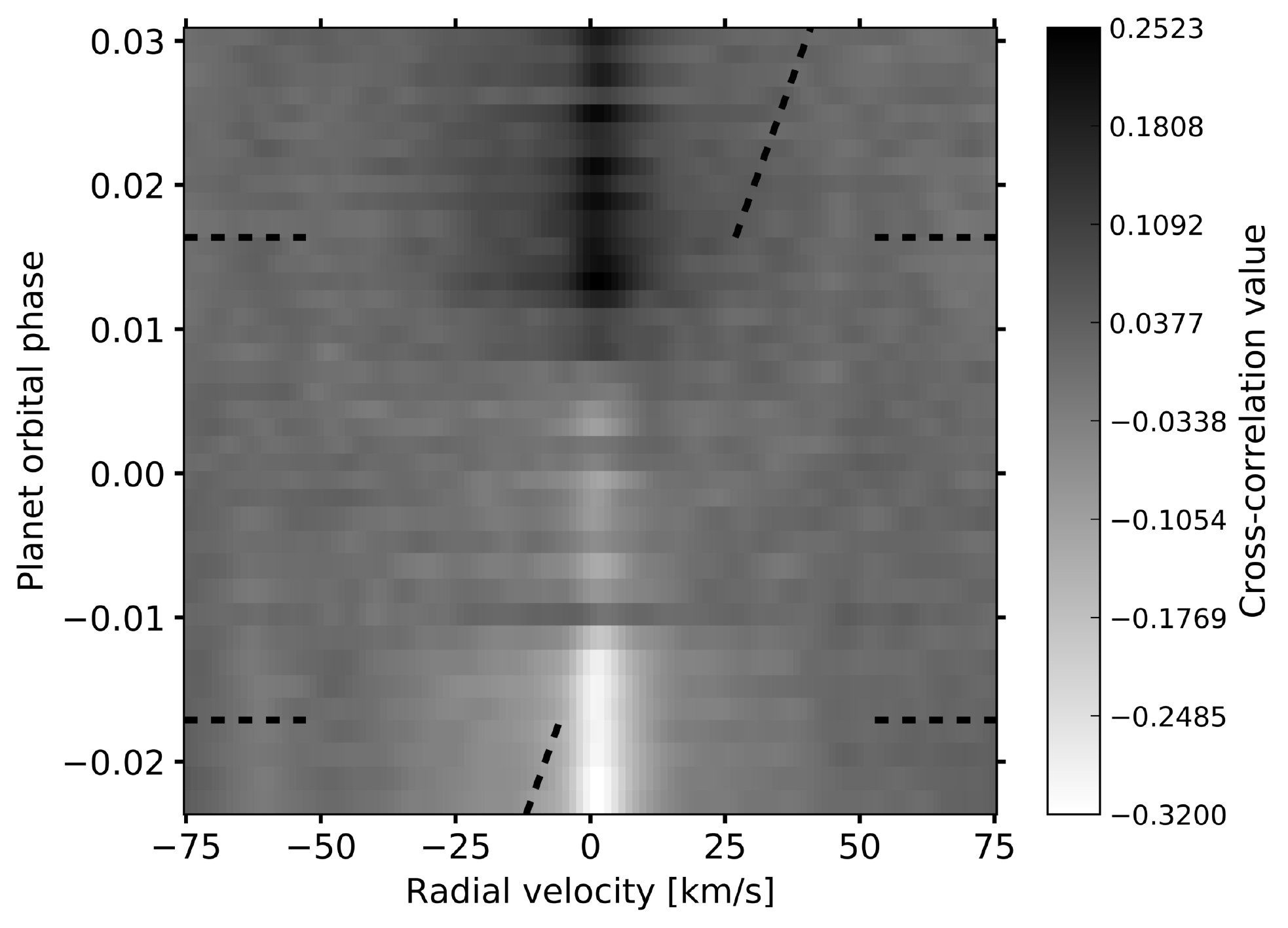}
\includegraphics[angle= 0, width=0.45\textwidth, height=0.22\textheight]{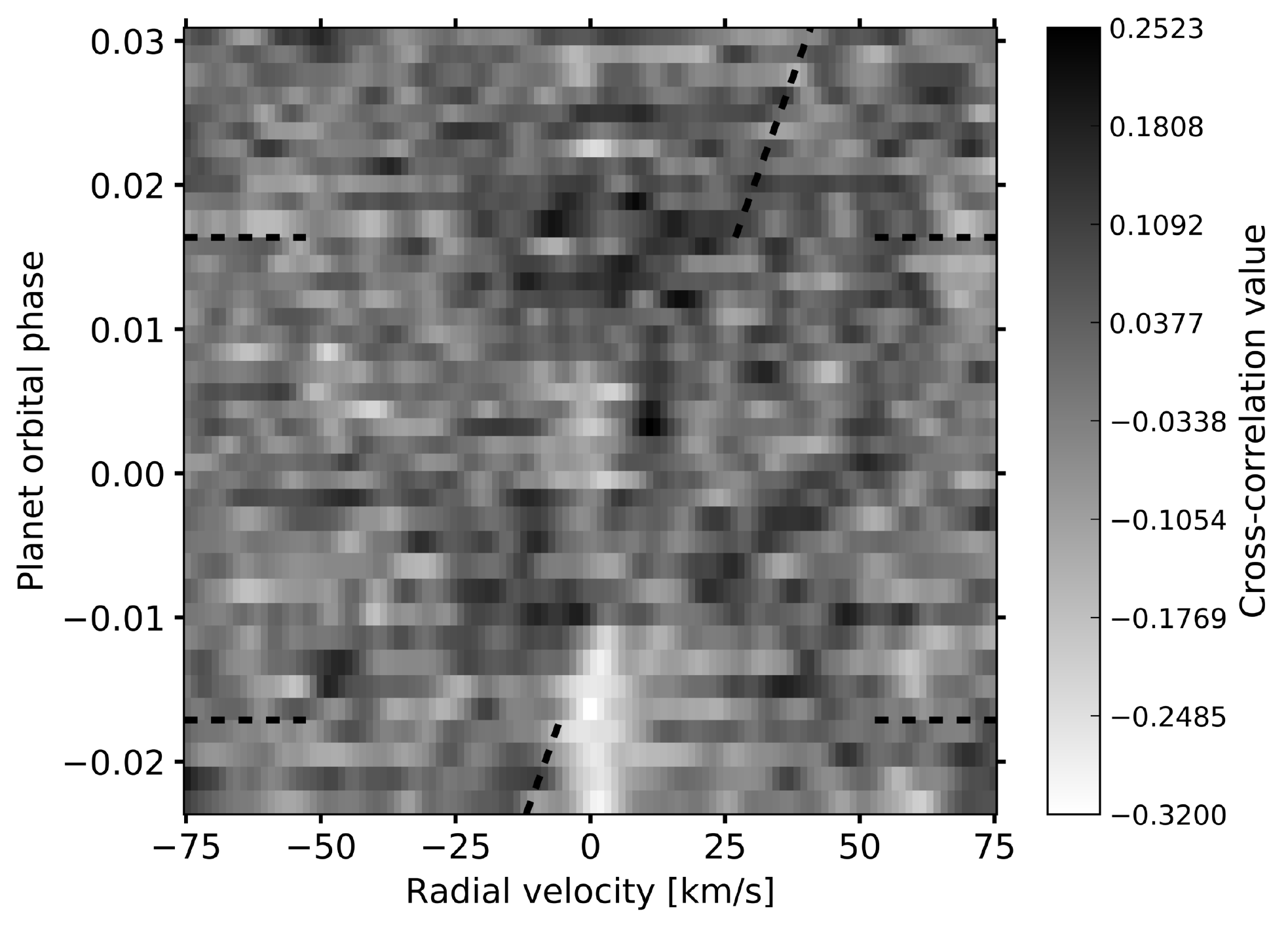}
\includegraphics[angle= 0, width=0.45\textwidth, height=0.22\textheight]{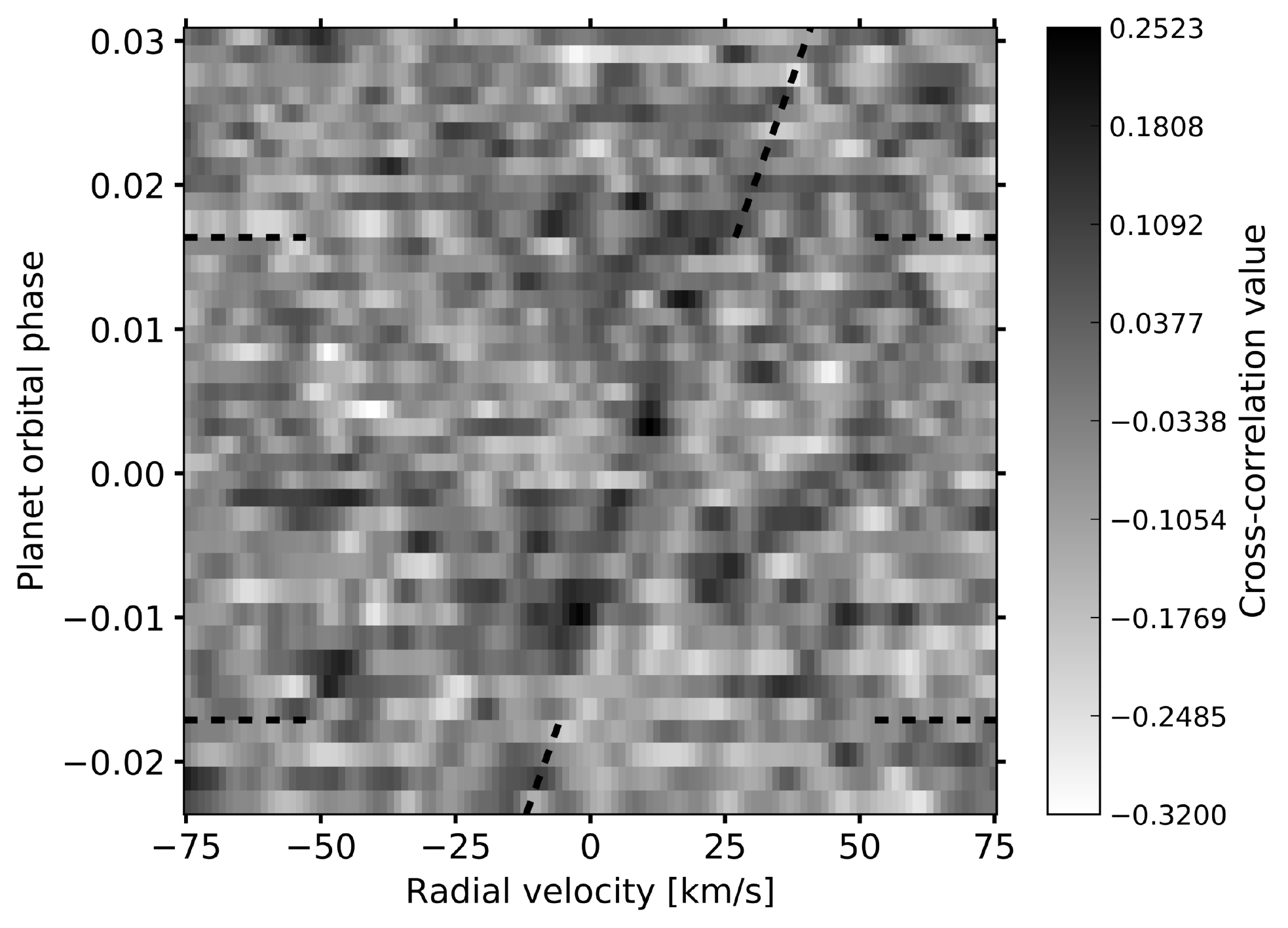}
\includegraphics[angle= 0, width=0.45\textwidth, height=0.22\textheight]{residuals_it10_6-24.pdf}
\caption{\label{fig.residuals}
Cross-correlation values as a function of the orbital phase and planet radial velocity with respect to the Earth after one ({\em top left}), two ({\em top right}), three ({\em bottom left}) and ten ({\em bottom right}) {\sc Sysrem} iterations. The first and last transit spectra are indicated with horizontal dashed lines. After the first and second iteration of {\sc Sysrem} the cross correlation with the telluric residuals are visible around the zero velocity. From the third iteration on, the suppression of the telluric residuals increases and the planet atmospheric signature is expected along the planetary velocities, indicated with the tilted dashed lines (when $\varv_{\rm wind}$\,=\,0\,km\,s$^{-1}$).
}
\end{figure*}

\subsection{Absorption spectral models}
\label{subsection.transmission_models}

We computed transmission spectra of HD\,189733\,b during its primary transit using the Karlsruhe Optimised and Precise Radiative Transfer Algorithm \citep[{\sc Kopra};][]{Stiller2002}. This code was originally developed for the Earth's atmosphere and was afterwards adapted to the atmospheres of Titan and Jupiter \citep{Garcia-Comas2011, Montanes2015}. {\sc Kopra} is a well-tested general purpose line-by-line radiative transfer model that includes all the known relevant processes for studying this problem.

In particular, our models include the spectral transitions of the molecules under study. 
We collected the molecular spectroscopic data from the HITEMP 2010 compilation \citep{Rothman2010} for H$_{2}$O and from the HITRAN 2012 compilation \citep{Rothman2013} for CH$_4$.
The line shapes were modelled with a Voigt profile. 
We used an adaptive scheme for including or rejecting spectral lines of a given strength (see \citealt{Stiller2002}), which is particularly useful for calculating the transmission of very hot planet atmospheres for which the line-list of species at high temperature contains a huge number of lines \citep{Rothman2010}. 

Collisions between molecular pairs of H$_2$-H$_2$ and H$_2$-He produce the so-called collision-induced absorption, resulting in ro-vibrational absorption bands. These bands are significant in a hot-Jupiter atmosphere spectrum, appearing mainly as smooth features in the 0.6--0.9\,$\mu$m, 1.0--1.4~$\,\mu$m, and 2.0--2.5\,$\mu$m spectral regions. We included these absorptions with the coefficients at high temperatures derived by \citet{Borysow2002} for H$_2$--H$_2$ pairs, and by \citet{Borysow1989} and \citet{Borysow1989a} 
for H$_2$--He. 
Once we computed the very high resolution ($\mathcal{R}$\,$\sim$\,4$\cdot$10$^{7}$) transmission spectra, we convolved them with the line spread function (LSF) of CARMENES.

We computed 12 synthetic transmission spectra. The first four models included only H$_{2}$O, another four models included only CH$_4$, and the last four models included both molecules. 

For the four water-only models, we used two different pressure-temperature profiles (A and B, see Fig.~\ref{fig.pt}) and two constant H$_{2}$O volume mixing ratios (VMRs) of 10$^{-5}$ and 10$^{-4}$.  
Although we did not expect significant differences on the results when using these models, due to the known model degeneracy \citep{Brogi14}, it was a robustness check of our detection. 
The $p$--$T$ profile A was obtained from our best fit (see Sect.\,\ref{subsection.bands}) to the transmission values measured with $HST$/WFC3 \citep{McCullough14}. 
The $p$--$T$ profile B was the one used by \cite{Brogi18}.
This $p$--$T$ profile is hotter than profile A in the lower atmosphere, below $\sim$$10^{-2}$\,bar (c.f. Fig.\,\ref{fig.pt}), where most of the absorption takes place.
Hence, the atmosphere of the $p$--$T$ B profile is more extended and, consequently, produces a larger absorption.  

The nominal transmission model used in the analysis was based on the $p$--$T$ profile B and a constant H$_{2}$O VMR of 10$^{-4}$ (see bottom panel of Fig.\,\ref{fig.spectra_and_model}).

We studied the presence of CH$_4$ in the planet atmosphere using eight models. All of them were calculated using the pressure--temperature profile B. Four of them included only methane, with CH$_4$ VMRs ranging from $10^{-7}$ to $10^{-4}$ in increments of one order of magnitude. The other four models included water, with a VMR of 10$^{-4}$, and methane with the previous range of VMRs.

During the process of removing the stellar and telluric features from the spectra  (Section\,\ref{subsection.telluric}), we removed also the continuum information. Therefore, before cross-correlating the residual matrices with the computed models, we also removed the continuum information from the models by subtracting their baseline level. 
Thus, the analysis is only sensitive to the strength of the absorption lines. 

\subsection{Cross correlation} \label{subsection.crosscorrelation}

The planet radial velocity changed during transit approximately from --10 to +25\,km\,s$^{-1}$  (see Fig.\,\ref{fig.residuals}). Thus, the cross-correlation analysis was performed for each spectral order over a wide range of planet radial velocities, from $-$130\,km\,s$^{-1}$ to +130\,km\,s$^{-1}$, in intervals of 1.3\,km\,s$^{-1}$ set by the mean velocity step-size of the CARMENES NIR pixels.
We linearly interpolated the molecular transmission templates to the corresponding Doppler-shifted wavelengths. 
The cross-correlation functions (CCFs) were obtained individually for each spectrum, forming a cross-correlation matrix ($\overline{CCF}$) per order. The dimension of each matrix was determined by the radial velocity lags used in the cross-correlation and the number of spectra, i.e., 201\,$\times$\,46.
The median value from each CCF was subtracted to account for possible broadband differences between the models and the spectra.
We applied the cross-correlation process after each {\sc Sysrem} iteration, obtaining 22 cross-correlation matrices, one per order used.

The cross-correlation matrices, $\overline{CCFs}$, coming from the same or different iterations, should be combined before retrieving the planet signal. In a previous work, \citet{Birkby17} injected synthetic signals in the data to find the optimal {\sc Sysrem} iteration for each of their four detectors and, subsequently, they combined the cross-correlation matrices of these different optimal iterations.  
However, this way of finding the optimal iteration depends on the model used and the strength of injection. 
\cite{Hawker2018} found similar issues when optimising the iteration choice by injecting the synthetic template.
This effect is more noticeable in our analysis than in that of \citet{Birkby17}, because the differences between models are more relevant the larger the studied wavelength range.
We followed \citet{Brogi18}, who minimised the model dependency of the analysis and equally co-added the $\overline{CCFs}$ that belonged to the same iteration.
A side effect of this choice is that it is possible to introduce correlation noise coming from over-corrected orders (those with the least signal) or spurious signals from under-corrected ones (those with the highest telluric contamination). 

Since we were most interested in studying the water bands fully covered by the CARMENES NIR channel, we chose to apply a first conservative analysis by using the useful orders from 1.06 to 1.58\,$\mu$m (from bluer to redder: orders 57, 56, 55, 52, 51, 50, 49, 48, 47, 46, 41, 40 and 39; see Fig.\,\ref{fig.spectra_and_model}). 
In this way, we optimised the iteration choice using the orders containing most of the signal (Sect.\,\ref{subsection.significance}), while possible contaminations by noise correlations of orders with the least expected signal were removed. Later, we included all the available orders (Sect.\,\ref{section.results}) and developed a multiband analysis of the data (Sect.\,\ref{subsection.bands}).

The cross-correlation matrices, for the first, second, third and tenth {\sc Sysrem} iterations are shown in Fig.~\ref{fig.residuals}. In the first two iterations (top panels), the cross-correlation functions are dominated by major telluric residuals around zero velocity, while for the third and tenth iterations (bottom panels), the main telluric and stellar components are removed. The atmospheric signal of the planet is expected to follow the planet radial velocity (i.e., the dashed slanted line in Fig.\,\ref{fig.residuals})  with positive cross-correlation values (in-transit).
Therefore, each spectrum needs to be shifted to the planet rest frame and combined before obtaining the total correlation signal.

The planet atmospheric signature moves through the in-transit spectra (i.e., from 21:01 UT to 23:11 UT) due to the change in its radial velocity ($\varv_{\rm P}$):
\begin{equation}
\label{equation.planet_velocity}
\varv_{\rm P}(t,K_{\rm P}) = \varv_{\rm sys} + \varv_{\rm bary}(t)  + K_{\rm P}\sin{2\pi\phi(t)},
\end{equation}
where $\varv_{\rm sys}$ is the systemic velocity of the stellar system, $\varv_{\rm bary}(t)$ is the barycentric velocity during observation,  $K_{\rm P}$ is the semi-amplitude of the planet radial velocity, and $\phi$$(t)$ is the planet orbital phase.

We shifted the cross-correlation function for each spectrum to the planet rest-frame using a linear interpolation of the velocities computed by Eq.\,\ref{equation.planet_velocity}. 
Subsequently, the planet trail was vertically aligned and all the in-transit CCFs were co-added to obtain a total CCF.
If a water signature is present, the peak of this CCF should be close to 0\,km\,s$^{-1}$.
However, the atmospheric dynamics of the planet can lead to non-zero values, as shown in previous works \citep{Louden15,Brogi16,Brogi18}.
We discuss this further in Sect.\,\ref{subsection.winds}.

We explored different $K_{\rm P}$ values to independently provide a direct measurement of the planet $K_{\rm P}$, which depends on the stellar mass (Table\,\ref{table.parameters}). In a similar way, we can probe the radial component of the high altitude global atmospheric winds ($\varv_{\rm wind}$) to obtain a better estimate of the noise properties.
We aligned the CCFs for a range of $K_{\rm P}$ values from 0 to 260\,km\,s$^{-1}$ and of $\varv_{\rm wind}$ values from --65 to +65\,km\,s$^{-1}$. For this, we included the additive term $\varv_{\rm wind}$ in Eq.\,\ref{equation.planet_velocity}.

We used this approach to study the significance of the retrieved signal and to verify that there are no significant spurious signals. In addition, the goodness of our telluric removal technique can be checked by exploring possible signals at low $K_{\rm P}$, i.e., in the Earth's rest frame.

\subsection{Significance of the retrieved signal}\label{subsection.significance}

\begin{figure}
\includegraphics[angle= 0, width=0.24\textwidth,trim=30 80 50 60,clip]{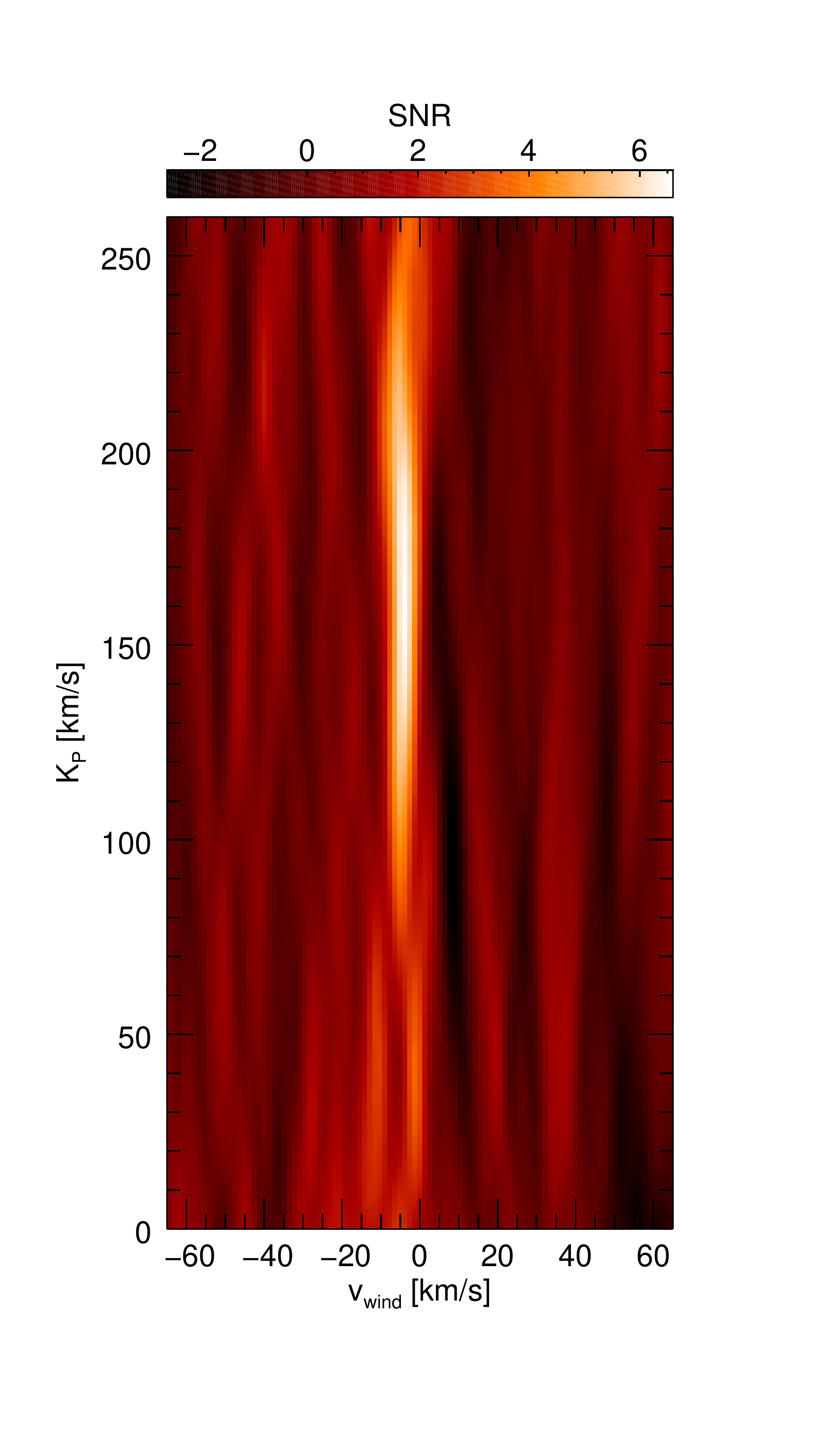}
\includegraphics[angle= 0, width=0.24\textwidth,trim=30 80 50 60,clip]{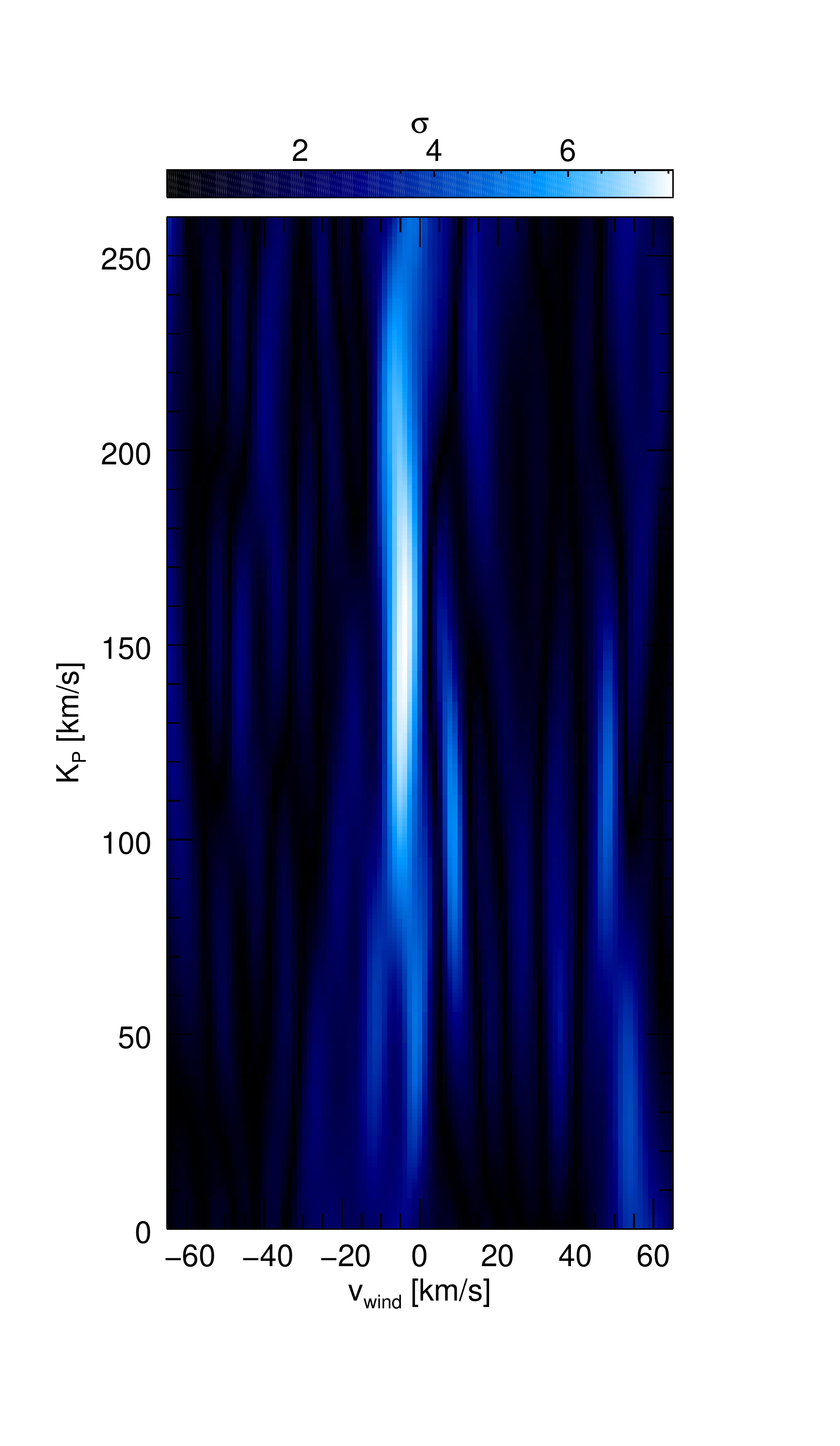}
\caption{\label{fig.significance}SNR results ({\em left}) and $t$-test results ({\em right}, in sigma units) when cross-correlating, in a wide range of $K_{\rm P}$ and $\varv_{\rm wind}$ values, the residual matrix and  the synthetic absorption model including the $p$--$T$ profile B and a water vapour content of VMR=10$^{-4}$. 
In both panels, the strongest signals reveal at planet radial velocity semi-amplitudes ($K_{\rm P}$) compatibles with values in the literature. They also happen at the same wind velocities ($\varv_{\rm wind}$), indicating a cross-correlation signal blue-shifted by --3.9$\pm$1.3\,km\,s$^{-1}$. No comparable signals are found close to the expected $K_{\rm P}$. There are neither comparable signals at the Earth's frame (low $K_{\rm P}$ values), discarding the presence of significant telluric contamination.}
\end{figure}

The significance of the retrieved signal was determined following similar methods to those of \citet{Birkby17} and \citet{Brogi18}. First, we computed the SNR by measuring the peak of the CCF and dividing it by its standard deviation excluding the signal.  
This provided a map of the SNR as a function of $K_{\rm P}$ and $\varv_{\rm wind}$, which revealed a pronounced signal whose position is consistent with a planetary origin (Fig.\,\ref{fig.significance}, left panel). 
However, this signal is spread over a wide range of $K_{\rm P}$ values. 
The change in the radial velocity during the transit is very small and does not depend enough on $K_{\rm P}$ to provide useful constrains.  
The uncertainties in $K_{\rm P}$ and $\varv_{\rm wind}$ are quoted as the limits where the SNR is one less than the peak value.

With a second method, we also used the in-transit spectra of the cross correlation matrix after alignment to the planet rest frame. The significance was obtained by performing a generalised $t$-test between the distribution of cross correlation values that do not carry signal (out-of-trail distribution) and the distribution of pixels that do carry signal (in-trail distribution). 

\begin{figure}
\includegraphics[angle= 0, width=0.49\textwidth]{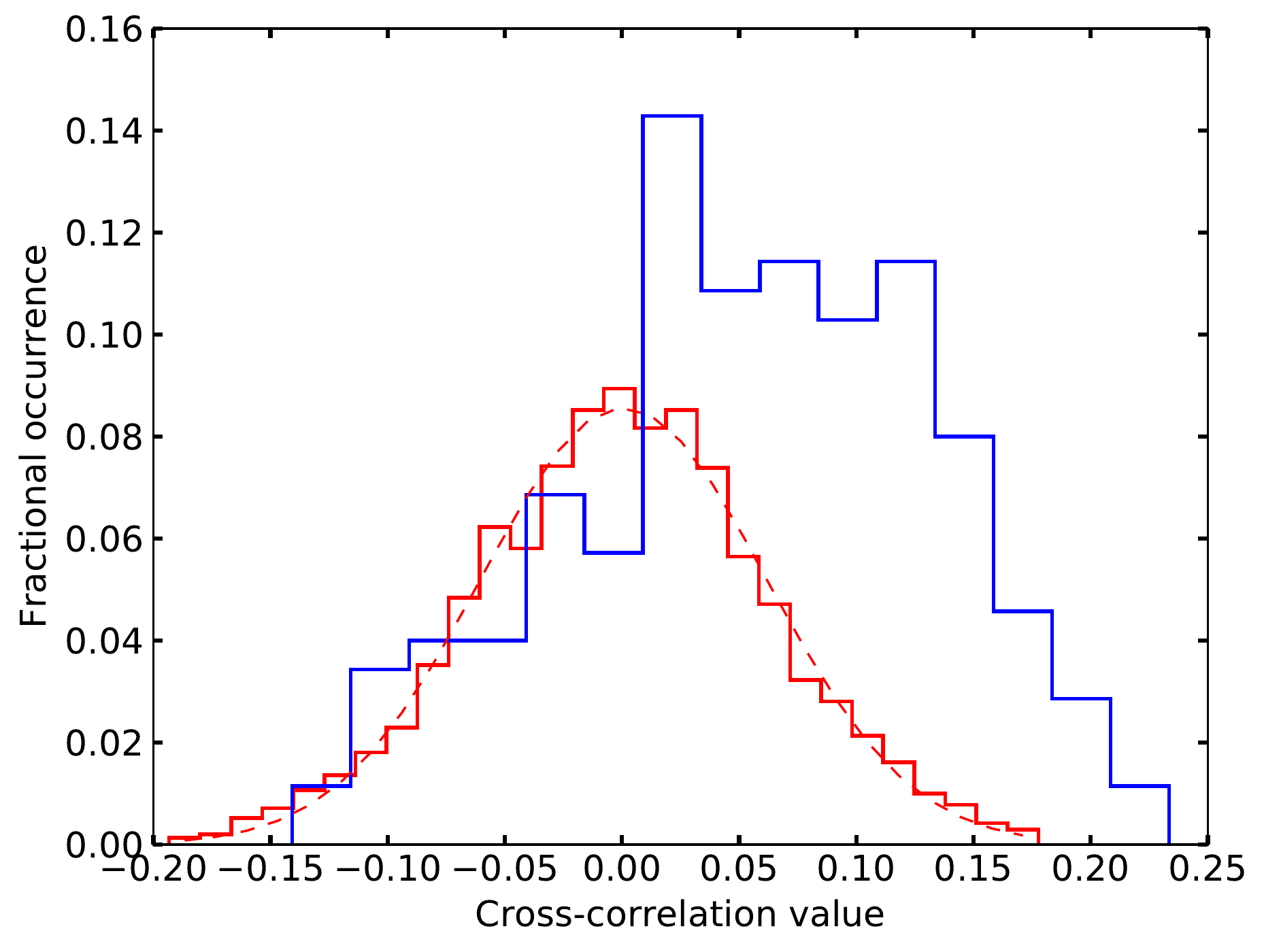}
\caption{\label{fig.distribution_comparison} Comparison between the distribution of cross-correlation values far away from the planet radial velocity (out-of-trail, red), a Gaussian distribution with the same mean and variance (dashed red line), and the distribution of the cross-correlation values near the planet radial velocities (in-trail, blue). The out-of-trail distribution follows the Gaussian distribution down to approximately 4$\sigma$. If the transmission signal of the planet were detected, both distributions would show significantly different means, as it is the case.}
\end{figure}

The out-of-trail distributions is composed by the pixels far away from the planet radial velocity  (i.e., all the pixels except for those within $\pm$15\,km\,s$^{-1}$, making a total of 3175 pixels). This distribution should be non-correlated noise and, therefore, should be normally distributed with a zero mean.
The in-trail distribution was defined as the values within $\pm$2.6\,km\,s$^{-1}$ around zero (i.e., 5 pixels width times 24 spectra) and its mean should be significantly different from zero in the case of a real planet signal.
The differences between the distributions can be seen in Fig.\,\ref{fig.distribution_comparison}, where we compare the two populations for the nominal transmission model at the optimal $\varv_{\rm wind}$ and $K_{\rm P}$ values. The out-of-trail population follows a normal distribution down to approximately 4\,$\sigma$.
The null hypothesis, H$_0$, states that both distributions have the same mean. 
The value of the $t$-statistic at which the hypothesis is rejected is then translated into a probability value (p-value) and expressed in terms of standard deviation $\sigma$.
The two populations present significantly different means and, therefore, the null hypothesis is discarded on a high level of significance ($\sigma$\,>\,7).
We applied this method for the same range of $\varv_{\rm wind}$ and $K_{\rm P}$ values as in the previous method. 
 
The results of the two methods,  when using the water absorption model computed with the $p$--$T$ profile B and a H$_{2}$O VMR\,=\,10$^{-4}$, are shown in Fig.\,\ref{fig.significance}.
The maximum significances in both methods are located at similar $K_{\rm P}$ and $\varv_{\rm wind}$ values, corresponding to the expected planet velocities. 

\begin{figure}
\includegraphics[angle= 0, width=0.49\textwidth, trim=20 0 25 35,clip]{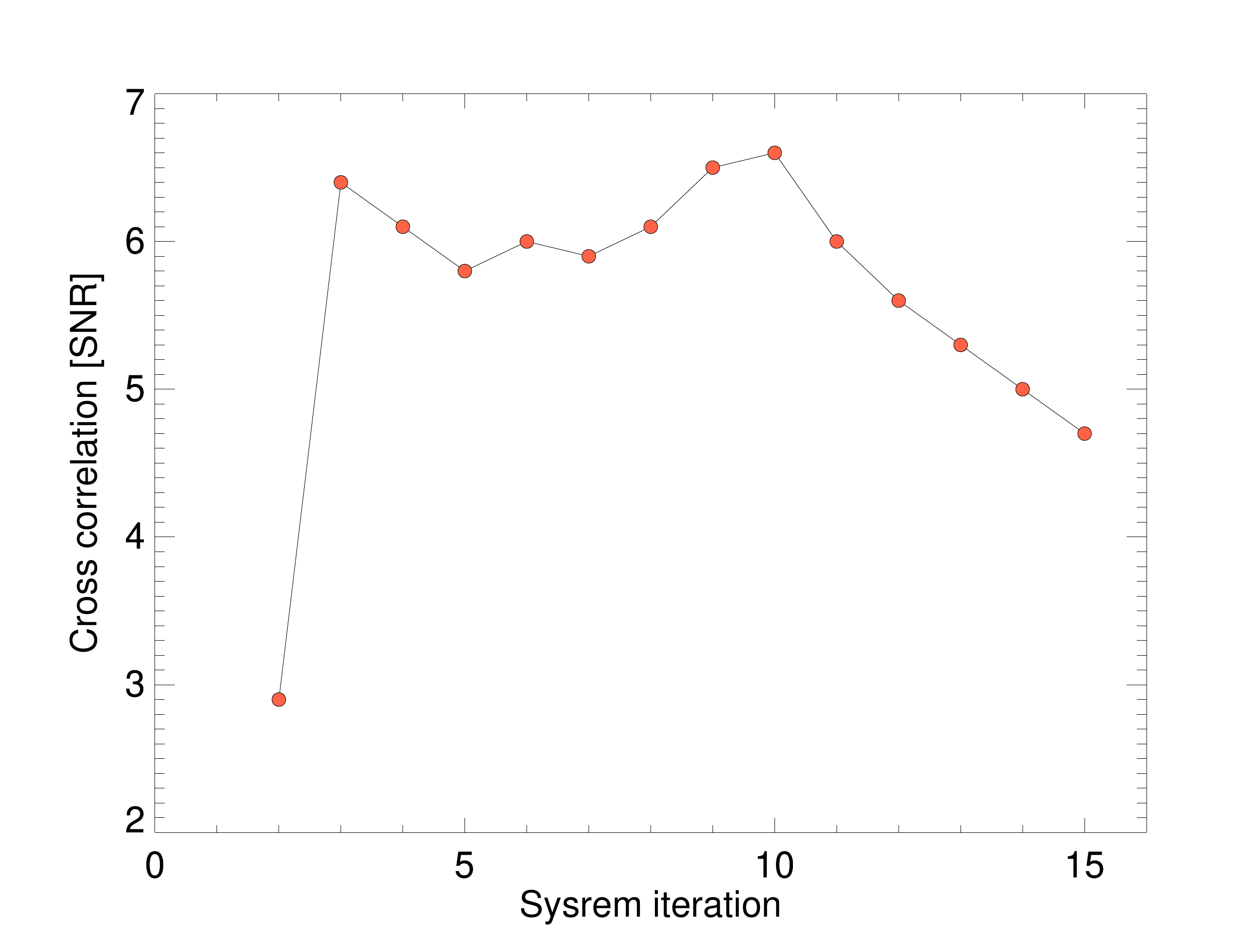}
\caption{\label{fig.iterations} The retrieved SNR when cross-correlating the residual matrix after each {\sc Sysrem} iteration with the H$_2$O absorption model including the $p$--$T$ profile B and a VMR\,=\,10$^{-4}$. The values correspond to the highest SNR at the corresponding $K_{\rm P}$ and $\varv_{\rm wind}$ in the range of $K_{\rm P}$\,=\,140 to 180\,km\,s$^{-1}$ and $\varv_{\rm wind}$=\,--10 to 0\,km\,s$^{-1}$. The first iteration is not shown for clarity, because it is totally dominated by telluric residuals.}
\end{figure}

We applied the SNR method to identify the iteration at which {\sc Sysrem} should be halted using the conservative approach explained in Sect.\,\ref{subsection.crosscorrelation}. The evolution of the maximum SNR recovered for the different {\sc Sysrem} iterations around the expected planet $K_{\rm P}$ using the nominal transmission model is shown in Fig.~\ref{fig.iterations}. 
The signal steadily increases in the first three iterations, as {\sc Sysrem} removes the major telluric and stellar contaminations.
After this point, the SNR is very similar until the 11$^{\rm th}$ iteration. 
Beyond this iteration, {\sc Sysrem} appears to mainly remove the exoplanet signal and, hence, the maximum SNR decreases.
We also noticed the decrease of the telluric signal observed at low $K_{\rm P}$ and  $\varv_{\rm wind}$  as the number of iteration increases (i.e., the contamination is smaller at the tenth than at the third iteration; see Fig.\,\ref{fig.residuals}). 
The behaviour of the SNR of the recovered CCF per iteration is similar for all the water absorption models tested in this work.

\citet{Hawker2018}, who detected the H$_{2}$O+CO band at 2.3\,$\mu$m and the HCN band at 3.2\,$\mu$m on HD\,209458\,b using {\sc Sysrem}, found that for most orders in their dataset a few {\sc Sysrem} iterations were enough to remove the contamination signals, while for heavily telluric contaminated ones, they required up to 13 iterations. 
We selected the tenth {\sc Sysrem} iteration for providing our results, as it is the iteration for which the results are most significant. 

\section{Results and discussion}\label{section.results}

\begin{figure}
\includegraphics[angle= 0, width=0.49\textwidth,trim=20 10 30 40,clip]{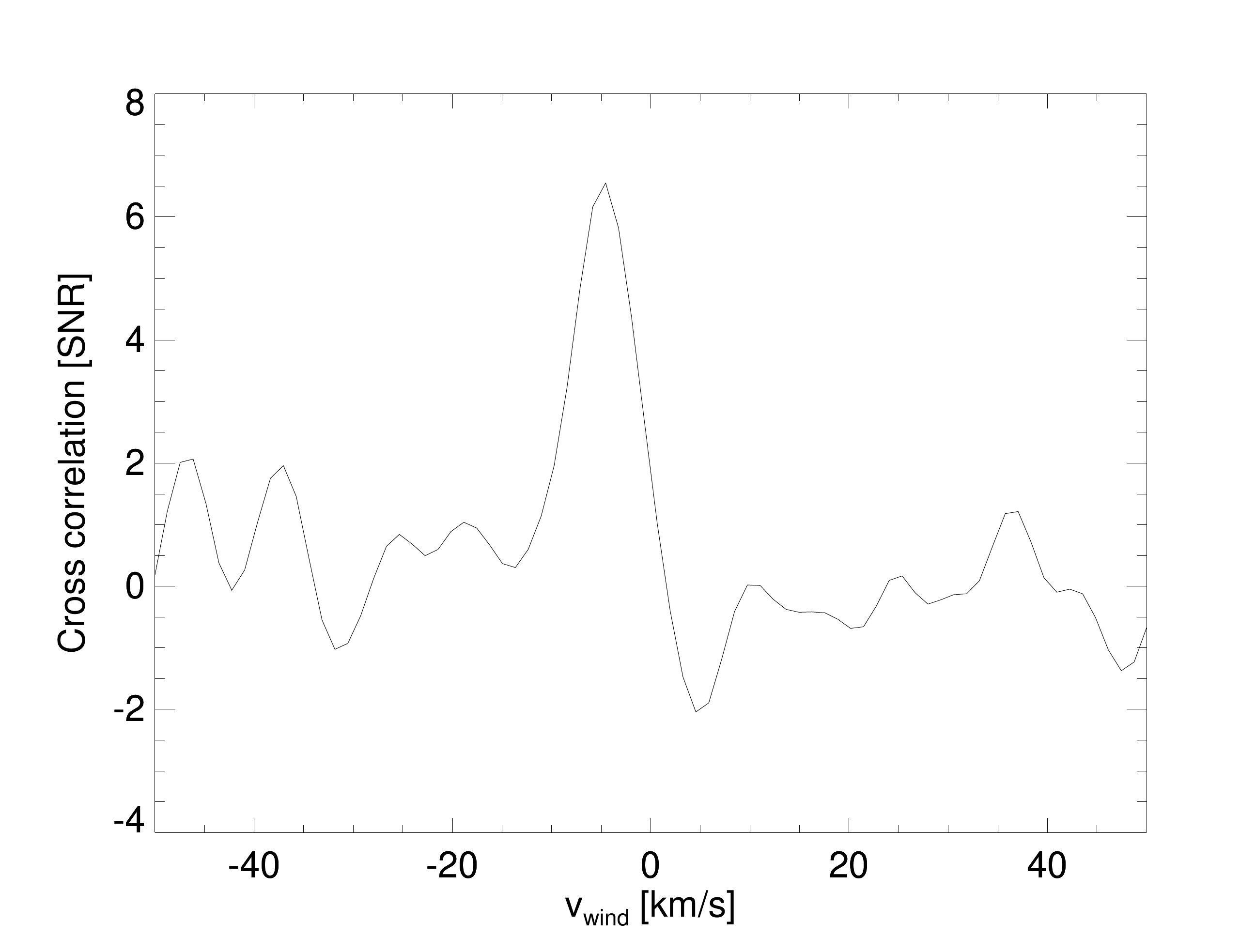}
\caption{\label{fig.bestCCF} Total CCF when using the nominal model and a conservative order selection (i.e., only the orders included in the two strongest bands). The signal is blue-shifted by --3.9$\pm$1.3\,km\,s$^{-1}$.}
\end{figure}

We retrieved the water vapour absorption signal of HD\,189733\,b in the CARMENES NIR data collected on 2017 September 7. 
When cross-correlating the residual matrix of this night using a conservative approach (including only useful orders from 1.06 to 1.58\,$\mu$m) with the absorption model calculated with a H$_2$O VMR of 10$^{-4}$ and the pressure--temperature profile B, we measured a maximum SNR\,=\,6.6 in the combined CCF (see Fig.\,\ref{fig.bestCCF}).
The generalised $t$-test analysis provides a p-value of 7.5\,$\sigma$ (see Fig.\,\ref{fig.significance}).
The results obtained with the absorption models computed with H$_2$O VMRs of 10$^{-5}$ and 10$^{-4}$ and the $p$-$T$ profiles A and B are similar and compatible within 1\,$\sigma$.

Independently of the used water model, we recovered the maximum SNR solution at compatible $K_{\rm P}$ and $\varv_{\rm wind}$ values.  
From the cross correlation with the nominal model, we obtained the maximum SNR at a planet radial velocity semi-amplitude of $K_{\rm P}$=160$^{+45}_{-33}$\,km\,s$^{-1}$. 
This value is compatible with previous results found in the literature as expected for the stellar mass (e.g., $K_{\rm P}$=152.5$^{+1.3}_{-1.8}$\,km\,s$^{-1}$; \citealt{Brogi16}).

When including all the useful orders in the CARMENES NIR wavelength range, we detected the atmospheric signature at similar significances levels as when using the conservative approximation (compare CCFs in Figs.\,\ref{fig.bestCCF} and \ref{fig.best_CCFs}). 
This was expected since the included orders carry a significantly smaller amount of water signal (see Sect.\,\ref{subsection.bands} for more information).

We also tested our analysis when injecting the negative of the templates into the observed spectra, before the telluric removal process, at $K_{\rm P}$\,=\,153\,km\,s$^{-1}$ and $\varv_{\rm wind}$\,=\,--3.9\,km\,s$^{-1}$. In this way, we found that the two models with water VMR\,=\,10$^{-5}$ were the best at cancelling the CCF signal, leaving a maximum residual of SNR$\sim$2. However, the two models with VMR\,=\,10$^{-4}$ over-cancelled the signal, leaving residuals of SNR$\sim$--3.5 (SNR\,<\,0 indicates negative values of the cross--correlation). These results agree with the values found by \citet{Madhusudhan2014}, who re-analysed the $HST$/WFC3 data from \citet{McCullough14} and constrained the water mixing ratio to be log(VMR)\,=\,--5.20$^{+1.68}_{-0.18}$. However, the presence of hazes in the atmosphere of HD\,189733\,b could provide similar depths of water absorption lines for higher water volume mixing ratios (e.g., \citealt{Madhusudhan2014,Pino18b}). 
Moreover, we did not detect the signal of the exoplanet atmosphere with any model including only methane at abundances from 10$^{-7}$ to 10$^{-4}$.
In fact, models including H$_2$O and CH$_4$ showed decreasing SNR with increasing CH$_4$ abundance. 
Hence, we did not find any evidence of methane in the atmosphere of HD\,189733\,b, confirming the results of \cite{Brogi18}.

\subsection{Winds in the atmosphere of HD\,189733\,b}\label{subsection.winds}
\begin{figure}
\includegraphics[angle= 0, width=0.49\textwidth]{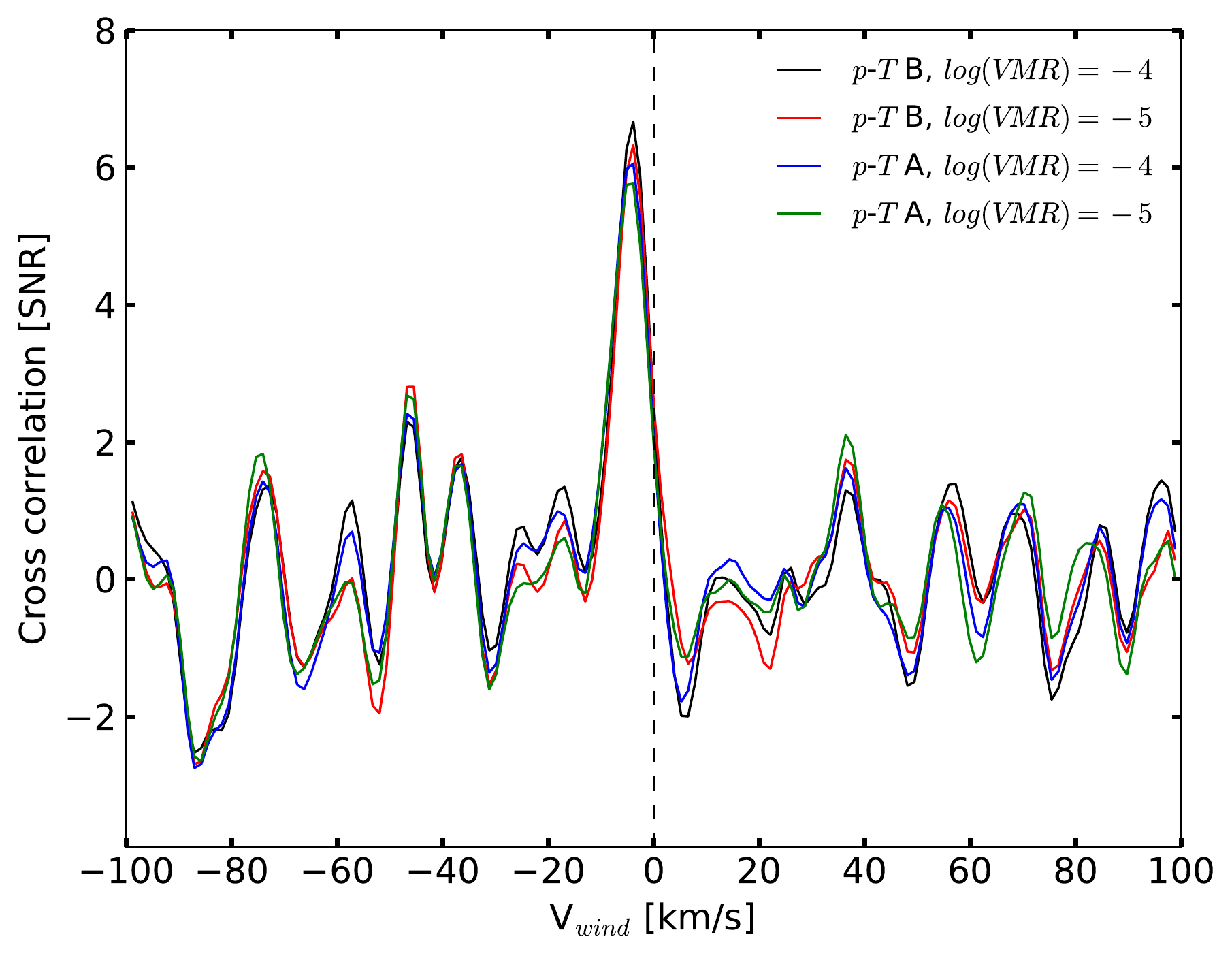}
\caption{\label{fig.best_CCFs}
Cross-correlation functions obtained when combining all the useful orders in the CARMENES NIR wavelength range for the studied H$_{2}$O absorption models. The results are very similar for the four tested models, hence showing a degeneracy in the tested $p$--$T$ profiles and H$_{2}$O abundances. 
}
\end{figure}

The total CCF (Figs.\,\ref{fig.bestCCF} and \ref{fig.best_CCFs}) shows a net blue--shifted water signal, $\varv_{\rm wind}$\,=\,--3.9\,$\pm$\,1.3\,km\,s$^{-1}$, which is comparable with previous measurements of sodium and water signals in HD\,189733\,b \citep{Louden15,Brogi16,Brogi18}.

The model of \citet{Showman2013} predicted that the atmosphere of HD\,189733\,b is in the transition between two extreme regimes of wind patterns being dominated either by strong zonal jets, or by global day- to night-side winds at the terminator. 
These atmospheric winds are blue- and red-shifted, and the combination of their signals during the whole transit is expected to leave a net blue-shifted wind signature on the atmospheric signal.

In particular, the application of the \citet{Showman2013} model predicts a blue-shifted velocity with a peak near $-$3\,km\,s$^{-1}$ for a fraction of $\sim$40\% of the terminator at a pressure level of 10$^{-4}$\,bar (see their Fig.\,7c).
Our derived value of --3.9\,$\pm$\,1.3\,km\,s$^{-1}$ agrees well with this prediction, although it is not fully comparable because it corresponds to an average of the whole terminator (i.e., blue- and red-shifted signals combined). 
Moreover, model contribution weighting functions (\citealt{Lee12,Blecic17}) predict that our signal, obtained mainly from the 1.15\,$\mu$m and 1.4\,$\mu$m water bands, should arise from atmospheric layers below the $\sim$10$^{-3}$\,bar level (i.e., P\,>\,10$^{-3}$\,bar).
In addition, the velocity value quoted above from \citet{Showman2013} did not include the planetary rotation. The inclusion of rotation would slightly increase the blue--shifted velocity (see their Fig.\,12). 

\begin{figure*}
\centering
\includegraphics[angle= 0, width=\textwidth, trim=40 20 40 20, clip]{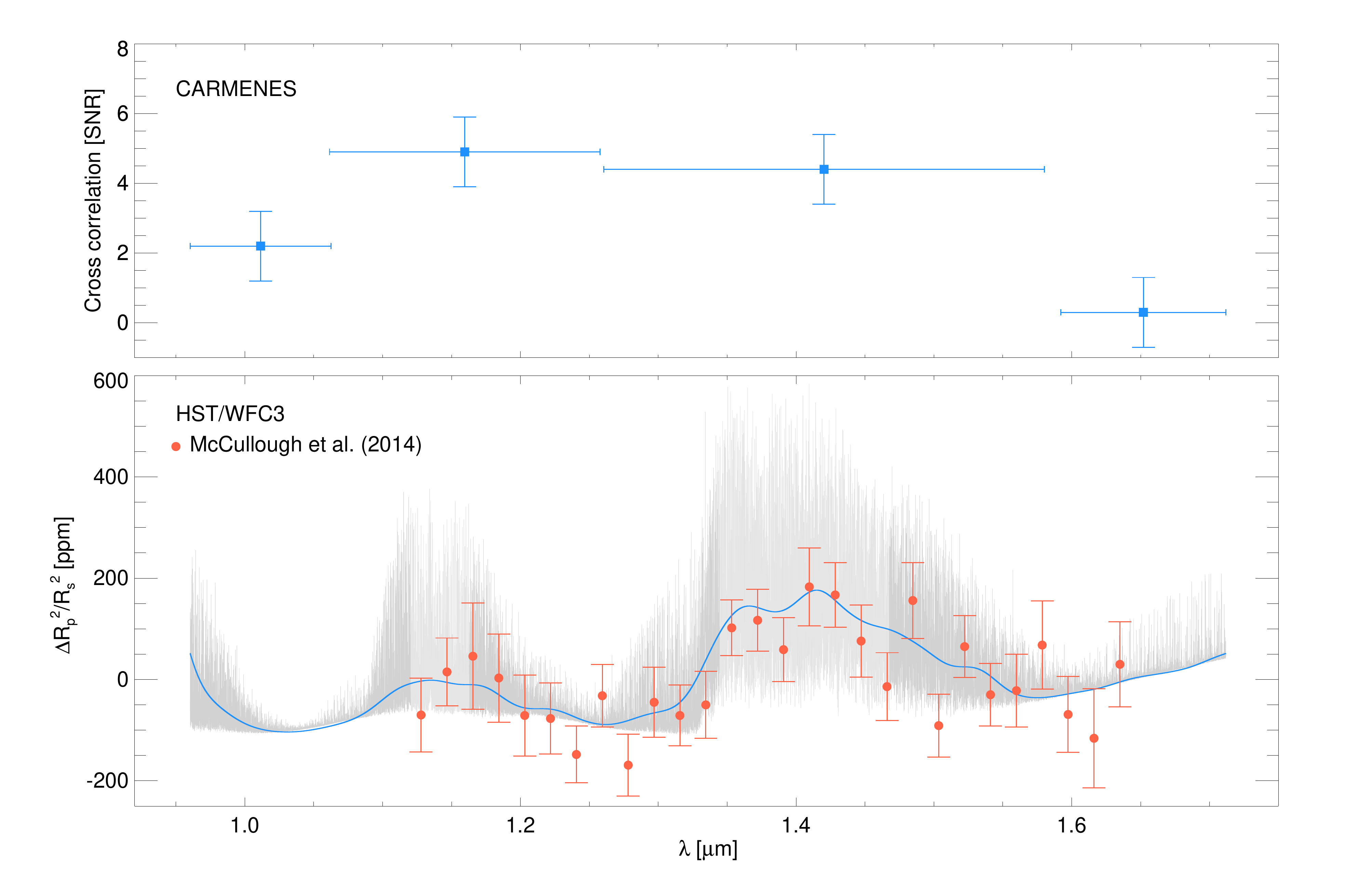}
\caption{\label{fig.bands} {\it Top panel}: SNR values of the CCFs for the four studied bands (see Table.\,\ref{table.bands}).
{\it Bottom panel}: water vapour absorption spectrum at the CARMENES resolution (in grey) when using the $p$--$T$ profile A and a water VMR\,=\,10$^{-5}$, and the same model but Gaussian smoothed to the $HST$/WFC3 resolution (in blue). The lighter grey areas are the wavelength regions covered by the orders discarded during our analysis.
Over-plotted in red are the $HST$/WFC3 measurements provided by \cite{McCullough14}.  We detected the two strongest bands around 1.15\,$\mu$m and 1.4\,$\mu$m. The reader should note that the SNRs shown in the top panel cannot be directly compared to the $HST$ absorption signals as shown in the bottom panel, since the former depend on the high frequency component of the transmission spectrum, and are strongly affected by telluric absorption.}
\end{figure*}

\subsection{Individual detections of the 1.15\,$\mu$m and 1.4\,$\mu$m H$_2$O bands}\label{subsection.bands}

We explored the possibility of individually detecting water vapour at different bands. We were particularly interested in detecting the two strongest bands at the wavelength ranges of 1.06--1.26\,$\mu$m (orders 58--50) and 1.26--1.58\,$\mu$m (orders 49--40).
For this analysis, we used the nominal transmission model and computed the SNR values at the optimal $K_{\rm P}$ and $\varv_{\rm wind}$ for four wavelength ranges.  The results are shown in Table\,\ref{table.bands} and plotted in Fig.\,\ref{fig.bands} (top panel). 
In order to better illustrate the spectral ranges used in the study of the different water bands, we have included (bottom panel, Fig.\,\ref{fig.bands}) the water transmission model that best match the data of \citet{McCullough14}. The maximum absorption reached in the high-resolution transmission spectra (in light grey) is $\sim$550\,p.p.m, which translates into $\sim$200\,p.p.m when convolving to the WFC3 resolution. This value agrees well with the $HST$ observations of \citet{McCullough14}.
We found H$_{2}$O signals for the strongest bands at similar signal-to-noise ratios, SNR$_{1.15}$\,=\,4.9 and SNR$_{1.4}$\,=\,4.4. 
Although the water band near 1.15\,$\mu$m is expected to be weaker than that at 1.4\,$\mu$m, a relevant number of orders were discarded during the analysis on the later, leaving similar maximum absorptions for both bands (see high-resolution transmission model in Fig.\,\ref{fig.bands}). This could explain the similarity between the retrieved SNR values. 

The bluest an reddest wavelength ranges do not provide a significant planet signal. The bluest orders (from 0.96 to 1.05\,$\mu$m), shown in Fig.\,\ref{fig.bands}, cover only a fraction of a water band, while hardly any water absorption is expected in the reddest orders. In addition, the first orders are affected by the reduced efficiency of the instrument (\citealt{Reiners18}).

Moreover, it is generally assumed that the 1.14 and 1.4\,$\mu$m water features, as well as those at shorter wavelengths, can be suppressed by the presence of hazes (e.g., \citealt{Sing16,Pino18a}). 
The recent work of \citet{Pino18b} suggested the use of the cross-correlation method in high-resolution transit spectra to detect water bands at optical and near-infrared wavelengths, which comparison could diagnose the presence of broad-band spectroscopic features, such as scattering by aerosols.
According to their work, the maximum CCF contrast (in the case of a clear atmosphere and no telluric contamination) expected between the bands studied here is of the order of 100\,p.p.m, i.e., $\sim$25\% of our deepest absorption lines. Therefore, it is not possible for us to derive any conclusion regarding the presence of aerosols in the atmosphere of HD\,189733\,b by comparing our multiple water band detections. 
However, the maximum CCF contrast between water features at shorter wavelengths and those studied here could be over 200\,p.p.m., which makes more feasible the detection of the hazes effect. 
Note that these numbers were derived using slightly different techniques and therefore should be taken with caution.

In a similar way, future instrumentation with a redder wavelength coverage than CARMENES, such as CRIRES+ \citep{Follert14}, could attempt the detection of water bands at K, L, and/or M band. These water bands are intrinsically stronger and the effects of hazes should be weaker than in J-band. \citet{Brogi16} detected water absorption in HD189733\,b using CRIRES at 2.3 $\mu$m, implying water vapour lines of up to $\sim$1.3$\times 10^{-3}$, about a factor two stronger than those at CARMENES wavelengths. This is in line with what is expected from the ratio in absorption strength. Uncertainties are currently too large to provide a constraint on the influence of hazes. 

\begin{table}
{\tiny
\centering
\caption{\label{table.bands} Signal-to-noise ratios and p-values (expressed in $\sigma$ values) of the CCFs for the four analysed wavelength ranges.}
\begin{tabular}{ccccc} 
   \hline
   \hline
   \noalign{\smallskip}
 $\Delta\lambda$ 					 & SNR & p-values  & $K_{\rm P}$  & $\varv_{\rm wind}$ \\
 
[$\mu$m] & & [$\sigma$]  & [km\,s$^{-1}$] & [km\,s$^{-1}$] \\ 
  \noalign{\smallskip}
    \hline
      \noalign{\smallskip}
0.96--1.06 & 2.2 & 3.8 & 153 & --3.9\\
 \noalign{\smallskip}
$^{a}$1.06--1.12, 1.16--1.26 & 4.9 & 6.8 & 146\,$\pm$\,46 & --3.9$^{+1.3}_{-2.6}$\\
  \noalign{\smallskip}
$^{a}$1.26--1.37, 1.47--1.58 & 4.4 & 5.7 & 156$^{+39}_{-31}$ & --3.9\,$\pm$\,1.3 \\ 
 \noalign{\smallskip}
 1.59--1.71 & 0.25 & 0.49 & 153 & --3.9\\
\noalign{\smallskip}
\hline
\end{tabular}
\tablefoot{The SNR and p-values were computed as in Sect.\,\ref{subsection.significance}. Since the water signals at the bluest and reddest bands are not detected, we used the fixed values of $K_{\rm P}$\,=\,153\,km\,s$^{-1}$ and $\varv_{\rm wind}$\,=\,--3.9\,km\,s$^{-1}$ to provide the results.   
\tablefoottext{a}{Band obtained by the combination of both wavelength ranges.}}
}
\end{table}

\section{Conclusions
\label{section.conclusions}}

We detected the near-infrared water signature on the atmosphere of HD\,189733\,b using CARMENES high-resolution spectra obtained during a single transit at SNR\,=\,6.6. The main problem for the detectability of H$_{2}$O from the ground is the telluric contamination.
However, an appropriate masking of the strongest telluric lines in combination with the use of {\sc Sysrem}, which performs a principal component analysis including proper error propagation, leads to an almost complete suppression of the contaminating signals even at relatively high precipitable water vapour levels.
On the contrary, we did not detect methane for any of the tested planet atmosphere models.

The water signal is detected at radial velocity semi-amplitude values compatible with the one of the planet. It is slightly more blue-shifted than previous works have measured, although compatible with them. Our value, --3.9\,$\pm$\,1.3\,km\,s$^{-1}$, is in good agreement with the results that \citet{Showman2013} derived using different dynamic atmospheric models for HD\,189733\,b. 
Follow-up observations aiming the detection of the water signal during the transit ingress and egress, for which our observations do not have enough signal, could help in deriving the blue- and red-shifted wind components and the circulation regime in the HD\,189733\,b middle atmosphere.

We exploited the large wavelength coverage of CARMENES to detect individually the water bands at 1.15 and 1.4\,$\mu$m, only conclusively observed before from the space with the $HST$/WFC3. 
The combination of the significances at which the individual water bands are detected agrees well with the overall SNR integrated over the whole CARMENES NIR wavelength range. This confirms that the wide instantaneous wavelength range of CARMENES significantly adds to the performance of the high-dispersion spectroscopy technique. 
Although, these water bands have been suggested for the study of hazes in the exoplanet atmospheres, the low contrast ratio expected between them \citep{Pino18b} do not allow us to derive any conclusion regarding the presence of hazes in the atmosphere of HD\,189733\,b. Whereby applicable with CARMENES, a simultaneous comparison between water bands at optical wavelengths and the ones presented here at higher signal-to-noise could provide an alternative way to constrain atmospheric hazes.
In addition, the different water bands can be included in multiband studies combining space and ground observations \citep{Brogi17}, capable to better constrain the chemical composition of the planet’s atmosphere.

In general, the results presented in this paper, the ones obtained with GIANO \citep{Brogi18}, the recent results of \citet{Yan18} with CARMENES and \citet{Hoeijmakers18} with HARPS, suggest that current and upcoming highly-stabilized high-resolution spectrographs ($\mathcal{R}$\,>\,20\,000) mounted on 4\,m-class telescopes will boost the study of exoplanet atmospheres from the ground.
 

\begin{acknowledgements}
We thank M. Brogi and the GIANO team for providing their $p-T$ profile and answering our questions regarding their work.
We thank J. Birkby for helpful discussions regarding the use of {\sc Sysrem}.
We also thank J. Hoeijmakers for the kind advises regarding the analysis of the data. 
F.J.A.-F. and I.S. acknowledge funding from the European Research Council (ERC) under the European Union Horizon 2020 research and innovation programme under grant agreement No 694513. 
CARMENES is funded by the German Max-Planck-Gesellschaft (MPG), the Spanish Consejo Superior de Investigaciones Cient\'ificas (CSIC), the European Union through European Regional Fund (FEDER/ERF), the Spanish Ministry of Economy and Competitiveness, the state of Baden-W\"urttemberg, the German Science Foundation (DFG), and the Junta de Andaluc\'ia, with additional contributions by the members of the CARMENES Consortium (Max-Planck-Institut f\"ur Astronomie, Instituto de Astrof\'isica de Andaluc\'ia, Landessternwarte K\"onigstuhl, Institut de Ci\`encies de l'Espai, Institut f\"ur Astrophysik G\"ottingen, Universidad Complutense de Madrid, Th\"uringer Landessternwarte Tautenburg, Instituto de Astrof\'isica de Canarias, Hamburger Sternwarte, Centro de Astrobiolog\'ia, and the Centro Astron\'omico Hispano-Alem\'an).
Financial support was also provided by the {Universidad Complutense de Madrid}, the Comunidad Aut\'onoma de Madrid, the Spanish Ministerios de Ciencia e Innovaci\'on and of Econom\'ia y Competitividad, the Fondo Europeo de Desarrollo Regional (FEDER/ERF), the Agencia estatal de investigaci{\'o}n, and the Fondo Social Europeo under grants 
ESP2014-54362-P, 
AYA2011-30147-C03-01, -02, and -03, 
AYA2012-39612-C03-01, 
ESP2013-48391-C4-1-R, 
ESP2014--54062--R, ESP 2016--76076--R, 
and BES--2015--074542. 
Based on observations collected at the Centro Astron\'omico Hispano Alem\'an (CAHA) at Calar Alto, operated jointly by the Max--Planck Institut f\"ur Astronomie and the Instituto de Astrof\'{\i}sica de Andaluc\'{\i}a. 
We thank the anonymous referee for their insightful comments, which contributed to improve the quality of the manuscript.
\end{acknowledgements}


\begin{thebibliography}{}
\bibitem[Agol et al.(2010)]{Agol10} Agol, E., Cowan, N.~B., Knutson, H.~A., et al.\ 2010, \apj, 721, 1861 
\bibitem[Birkby et al.(2013)]{Birkby13} Birkby, J.~L., de Kok, R.~J., Brogi, M., et al.\ 2013, \mnras, 436, L35 
\bibitem[Birkby et al.(2017)]{Birkby17} Birkby, J.~L., de Kok, R.~J., Brogi, M., Schwarz, H., \& Snellen, I.~A.~G.\ 2017, \aj, 153, 138 
\bibitem[Blecic et al.(2017)]{Blecic17} Blecic, J., Dobbs-Dixon, I., \& Greene, T.\ 2017, \apj, 848, 127 
\bibitem[{Borysow(2002)}]{Borysow2002} Borysow, A. 2002, \aap, 390, 779
\bibitem[{{Borysow} \& {Frommhold}(1989)}]{Borysow1989a}
{Borysow}, A. \& {Frommhold}, L. 1989, \apj, 341, 549
\bibitem[{Borysow {et~al.}(1989)Borysow, Frommhold, \& Moraldi}]{Borysow1989}
Borysow, A., Frommhold, L., \& Moraldi, M. 1989, \apj, 336,
  495
\bibitem[Bouchy et al.(2005)]{Bouchy05} Bouchy, F., Udry, S., Mayor, M., et al.\ 2005, \aap, 444, L15 
\bibitem[Bourrier et al.(2013)]{Bourrier13} Bourrier, V., Lecavelier des Etangs, A., Dupuy, H., et al.\ 2013, \aap, 551, A63 
\bibitem[Brogi et al.(2012)]{Brogi12} Brogi, M., Snellen, I.~A.~G., de Kok, R.~J., et al.\ 2012, \nat, 486, 502 
\bibitem[Brogi et al.(2013)]{Brogi13} Brogi, M., Snellen, I.~A.~G., de Kok, R.~J., et al.\ 2013, \apj, 767, 27 
\bibitem[Brogi et al.(2014)]{Brogi14} Brogi, M., de Kok, R.~J., Birkby, J.~L., Schwarz, H., \& Snellen, I.~A.~G.\ 2014, \aap, 565, A124 
\bibitem[Brogi et al.(2016)]{Brogi16} Brogi, M., de Kok, R.~J., Albrecht, S., et al.\ 2016, \apj, 817, 106
\bibitem[Brogi et al.(2017)]{Brogi17} Brogi, M., Line, M., Bean, J., D{\'e}sert, J.-M., \& Schwarz, H.\ 2017, \apjl, 839, L2 
\bibitem[Brogi et al.(2018)]{Brogi18} Brogi, M., Giacobbe, P., Guilluy, G., et al.\ 2018, \aap, 615, A16 
\bibitem[Brown(2001)]{Brown01} Brown, T.~M.\ 2001, \apj, 553, 1006 
\bibitem[Caballero et al.(2016)]{Caballero2016} Caballero, J.~A., Gu{\`a}rdia, J., L{\'o}pez del Fresno, M., et al.\ 2016, \procspie, 9910, 99100E 
\bibitem[Charbonneau et al.(1999)]{Charbonneau99} Charbonneau, D., Noyes, R.~W., Korzennik, S.~G., et al.\ 1999, \apjl, 522, L145 
\bibitem[Charbonneau et al.(2002)]{Charbonneau02} Charbonneau, D., Brown, T.~M., Noyes, R.~W., \& Gilliland, R.~L.\ 2002, \apj, 568, 377 
\bibitem[Charbonneau et al.(2005)]{Charbonneau05} Charbonneau, D., Allen, L.~E., Megeath, S.~T., et al.\ 2005, \apj, 626, 523 
\bibitem[Charbonneau et al.(2008)]{Charbonneau08} Charbonneau, D., Knutson, H.~A., Barman, T., et al.\ 2008, \apj, 686, 1341 
\bibitem[Collier Cameron et al.(1999)]{Collier-Cameron99} Collier Cameron, A., Horne, K., Penny, A., \& James, D.\ 1999, \nat, 402, 751 
\bibitem[Deming et al.(2005a)]{Deming05a} Deming, D., Brown, T.~M., Charbonneau, D., Harrington, J., \& Richardson, L.~J.\ 2005, \apj, 622, 1149 
\bibitem[Deming et al.(2005b)]{Deming05b} Deming, D., Seager, S., Richardson, L.~J., \& Harrington, J.\ 2005, \nat, 434, 740 
\bibitem[Deming et al.(2006)]{Deming06} Deming, D., Harrington, J., Seager, S., \& Richardson, L.~J.\ 2006, \apj, 644, 560 
\bibitem[de Kok et al.(2013)]{deKok13} de Kok, R.~J., Brogi, M., Snellen, I.~A.~G., et al.\ 2013, \aap, 554, A82 
\bibitem[Follert et al.(2014)]{Follert14} Follert, R., Dorn, R.~J., Oliva, E., et al.\ 2014, \procspie, 9147, 914719 
\bibitem[Gaia Collaboration et al.(2018)]{Gaia18} Gaia Collaboration, Brown, A.~G.~A., Vallenari, A., et al.\ 2018, \aap, 616, A1 
\bibitem[{Garc{\'i}a-Comas {et~al.}(2011)Garc{\'i}a-Comas, L{\'o}pez-Puertas,
  Funke, Dinelli, Moriconi, Adriani, Molina, \& Coradini}]{Garcia-Comas2011}
Garc{\'i}a-Comas, M., L{\'o}pez-Puertas, M., Funke, B., {et~al.} 2011, Icarus,
  214, 571
\bibitem[Gibson et al.(2012)]{Gibson12} Gibson, N.~P., Aigrain, S., Pont, F., et al.\ 2012, \mnras, 422, 753 
\bibitem[Grillmair et al.(2008)]{Grillmair08} Grillmair, C.~J., Burrows, A., Charbonneau, D., et al.\ 2008, \nat, 456, 767 
\bibitem[Hawker et al.(2018)]{Hawker2018} Hawker, G.~A., Madhusudhan, N., Cabot, S.~H.~C., \& Gandhi, S.\ 2018, \apjl, 863, L11 
\bibitem[Heng \& Kitzmann(2017)]{Heng2017} Heng, K., \& Kitzmann, D.\ 2017, \mnras, 470, 2972 
\bibitem[Hoeijmakers et al.(2018)]{Hoeijmakers18} Hoeijmakers, H.~J., Ehrenreich, D., Heng, K., et al.\ 2018, \nat, 560, 453 
\bibitem[Hubbard et al.(2001)]{Hubbard01} Hubbard, W.~B., Fortney, J.~J., Lunine, J.~I., et al.\ 2001, \apj, 560, 413 
\bibitem[Kaeufl et al.(2004)]{Kaeufl04} Kaeufl, H.-U., Ballester, P., Biereichel, P., et al.\ 2004, \procspie, 5492, 1218 
\bibitem[\protect\citeauthoryear{Koen et al.}{2010}]{Koen10} Koen, C., Kilkenny, D., van Wyk, F., Marang, F., 2010, MNRAS, 403, 1949
\bibitem[Lecavelier Des Etangs et al.(2008)]{Lecavelier08} Lecavelier Des Etangs, A., Pont, F., Vidal-Madjar, A., \& Sing, D.\ 2008, \aap, 481, L83 
\bibitem[Lecavelier Des Etangs et al.(2010)]{Lecavelier10} Lecavelier Des Etangs, A., Ehrenreich, D., Vidal-Madjar, A., et al.\ 2010, \aap, 514, A72 
\bibitem[Lecavelier des Etangs et al.(2012)]{Lecavelier12} Lecavelier des Etangs, A., Bourrier, V., Wheatley, P.~J., et al.\ 2012, \aap, 543, L4 
\bibitem[Lee et al.(2012)]{Lee12} Lee, J.-M., Fletcher, L.~N., \& Irwin, P.~G.~J.\ 2012, \mnras, 420, 170 
\bibitem[Lockwood et al.(2014)]{Lockwood14} Lockwood, A.~C., Johnson, J.~A., Bender, C.~F., et al.\ 2014, \apjl, 783, L29 
\bibitem[Louden \& Wheatley(2015)]{Louden15} Louden, T., \& Wheatley, P.~J.\ 2015, \apjl, 814, L24 
\bibitem[Madhusudhan et al.(2014)]{Madhusudhan2014} Madhusudhan, N., Crouzet, N., McCullough, P.~R., Deming, D., \& Hedges, C.\ 2014, \apjl, 791, L9 
\bibitem[Mazeh et al.(2007)]{Mazeh07} Mazeh, T., Tamuz, O., \& Zucker, S.\ 2007, Transiting Extrasolar Planets Workshop, 366, 119 
\bibitem[Mayor \& Queloz(1995)]{Mayor95} Mayor, M., \& Queloz, D.\ 1995, \nat, 378, 355 
\bibitem[Mayor et al.(2003)]{Mayor03} Mayor, M., Pepe, F., Queloz, D., et al.\ 2003, The Messenger, 114, 20 
\bibitem[McCullough et al.(2014)]{McCullough14} McCullough, P.~R., Crouzet, N., Deming, D., \& Madhusudhan, N.\ 2014, \apj, 791, 55 
\bibitem[{Monta{\~n}{\'e}s-Rodr{\'\i}guez
  {et~al.}(2015)Monta{\~n}{\'e}s-Rodr{\'\i}guez, Gonz{\'a}lez-Merino, Palle,
  L{\'o}pez-Puertas, \& Garc{\'\i}a-Melendo}]{Montanes2015}
Monta{\~n}{\'e}s-Rodr{\'\i}guez, P., Gonz{\'a}lez-Merino, B., Palle, E.,
  L{\'o}pez-Puertas, M., \& Garc{\'\i}a-Melendo, E. 2015, {\apj}L,
  801, L8
\bibitem[Nugroho et al.(2017)]{Nugroho17} Nugroho, S.~K., Kawahara, H., Masuda, K., et al.\ 2017, \aj, 154, 221 
\bibitem[Pino et al.(2018a)]{Pino18a} Pino, L., Ehrenreich, D., Wyttenbach, A., et al.\ 2018, \aap, 612, A53 
\bibitem[Pino et al.(2018b)]{Pino18b} Pino, L., Ehrenreich, D., Allart, R., et al.\ 2018, arXiv:1807.10769 
\bibitem[Piskorz et al.(2016)]{Piskorz16} Piskorz, D., Benneke, B., Crockett, N.~R., et al.\ 2016, \apj, 832, 131 
\bibitem[Piskorz et al.(2017)]{Piskorz17} Piskorz, D., Benneke, B., Crockett, N.~R., et al.\ 2017, \aj, 154, 78 
\bibitem[Pont et al.(2008)]{Pont08} Pont, F., Knutson, H., Gilliland, R.~L., Moutou, C., \& Charbonneau, D.\ 2008, \mnras, 385, 109 
\bibitem[Pont et al.(2013)]{Pont13} Pont, F., Sing, D.~K., Gibson, N.~P., et al.\ 2013, \mnras, 432, 2917 
\bibitem[Quirrenbach et al.(2016)]{Quirrenbach16} Quirrenbach, A., Amado, P.~J., Caballero, J.~A., et al.\ 2016, \procspie, 9908, 990812 
\bibitem[Redfield et al.(2008)]{Redfield08} Redfield, S., Endl, M., Cochran, W.~D., \& Koesterke, L.\ 2008, \apjl, 673, L87 
\bibitem[Reiners et al.(2018)]{Reiners18} Reiners, A., Zechmeister, M., Caballero, J.~A., et al.\ 2018, \aap, 612, A49 
\bibitem[Rodler et al.(2008)]{Rodler08} Rodler, F., K{\"u}rster, M., \& Henning, T.\ 2008, \aap, 485, 859 
\bibitem[Rodler et al.(2010)]{Rodler10} Rodler, F., K{\"u}rster, M., \& Henning, T.\ 2010, \aap, 514, A23 
\bibitem[Rodler et al.(2012)]{Rodler12} Rodler, F., L{\'o}pez-Morales, M., \& Ribas, I.\ 2012, \apjl, 753, L25 
\bibitem[{Rothman {et~al.}(2010)Rothman, Gordon, Barber, Dothe, Gamache,
  Goldman, Perevalov, Tashkun, \& Tennyson}]{Rothman2010}
Rothman, L.~S., Gordon, I.~E., Barber, R.~J., {et~al.} 2010, Journal of
  Quantitative Spectroscopy and Radiative Transfer, 111, 2139
\bibitem[{Rothman {et~al.}(2013)Rothman, Gordon, Babikov, Barbe, Chris~Benner,
 Bernath, Birk, Bizzocchi, Boudon, Brown, Campargue, Chance, Cohen, Coudert,
 Devi, Drouin, Fayt, Flaud, Gamache, Harrison, Hartmann, Hill, Hodges,
 Jacquemart, Jolly, Lamouroux, Le~Roy, Li, Long, Lyulin, Mackie, Massie,
 Mikhailenko, M{\"u}ller, Naumenko, Nikitin, Orphal, Perevalov, Perrin,
 Polovtseva, Richard, Smith, Starikova, Sung, Tashkun, Tennyson, Toon,
 Tyuterev, \& Wagner}]{Rothman2013}
Rothman, L.~S., Gordon, I.~E., Babikov, Y., {et~al.} 2013, Journal of
 Quantitative Spectroscopy and Radiative Transfer, 130, 4
\bibitem[Schwarz et al.(2015)]{Schwarz15} Schwarz, H., Brogi, M., de Kok, R., Birkby, J., \& Snellen, I.\ 2015, \aap, 576, A111 
\bibitem[Schwarz et al.(2016)]{Schwarz16} Schwarz, H., Ginski, C., de Kok, R.~J., et al.\ 2016, \aap, 593, A74 
\bibitem[Seager \& Sasselov(2000)]{Seager00} Seager, S., \& Sasselov, D.~D.\ 2000, \apj, 537, 916 
\bibitem[Seifert et al.(2012)]{Seifert12} Seifert, W., S{\'a}nchez Carrasco, M.~A., Xu, W., et al.\ 2012, \procspie, 8446, 844633 
\bibitem[{Showman {et~al.}(2013)Showman, Fortney, Lewis, \&
  Shabram}]{Showman2013}
Showman, A.~P., Fortney, J.~J., Lewis, N.~K., \& Shabram, M. 2013, The
  \apj, 762, 1
\bibitem[Sing et al.(2011)]{Sing11} Sing, D.~K., Pont, F., Aigrain, S., et al.\ 2011, \mnras, 416, 1443 
\bibitem[Sing et al.(2016)]{Sing16} Sing, D.~K., Fortney, J.~J., Nikolov, N., et al.\ 2016, \nat, 529, 59 
\bibitem[Skrutskie et al.(2006)]{Skrutskie06} Skrutskie, M.~F., Cutri, R.~M., Stiening, R., et al.\ 2006, \aj, 131, 1163 
\bibitem[{Smette {et~al.}(2015)Smette, Sana, Noll, Horst, Kausch, Kimeswenger,
  Barden, Szyszka, Jones, Gallenne, Vinther, Ballester, \&
  Taylor}]{Smette2015}
Smette, A., Sana, H., Noll, S., {et~al.} 2015, Astronomy and Astrophysics, 576,
  A77
\bibitem[Snellen et al.(2008)]{Snellen08} Snellen, I.~A.~G., Albrecht, S., de Mooij, E.~J.~W., \& Le Poole, R.~S.\ 2008, \aap, 487, 357 
\bibitem[Snellen et al.(2010)]{Snellen10} Snellen, I.~A.~G., de Kok, R.~J., de Mooij, E.~J.~W., \& Albrecht, S.\ 2010, \nat, 465, 1049 
\bibitem[Stevenson(2016)]{Stevenson2016} Stevenson, K.~B.\ 2016, \apjl, 817, L16 
\bibitem[{Stiller {et~al.}(2002)Stiller, von Clarmann, Funke, Glatthor, Hase,
  H{\"o}pfner, \& Linden}]{Stiller2002}
Stiller, G.~P., von Clarmann, T., Funke, B., {et~al.} 2002, Journal of
  Quantitative Spectroscopy and Radiative Transfer, 72, 249
\bibitem[St{\"u}rmer et al.(2014)]{Sturmer14} St{\"u}rmer, J., Stahl, O., Schwab, C., et al.\ 2014, \procspie, 9151, 915152 
 \bibitem[Swain et al.(2010)]{Swain10} Swain, M.~R., Deroo, P., Griffith, C.~A., et al.\ 2010, \nat, 463, 637 
\bibitem[Tamuz et al.(2005)]{Tamuz05} Tamuz, O., Mazeh, T., \& Zucker, S.\ 2005, \mnras, 356, 1466 
\bibitem[Todorov et al.(2014)]{Todorov14} Todorov, K.~O., Deming, D., Burrows, A., \& Grillmair, C.~J.\ 2014, \apj, 796, 100 
\bibitem[Torres et al.(2008)]{Torres08} Torres, G., Winn, J.~N., \& Holman, M.~J.\ 2008, \apj, 677, 1324-1342 
\bibitem[Triaud et al.(2009)]{Triaud09} Triaud, A.~H.~M.~J., Queloz, D., Bouchy, F., et al.\ 2009, \aap, 506, 377 
\bibitem[Wyttenbach et al.(2015)]{Wyttenbac15} Wyttenbach, A., Ehrenreich, D., Lovis, C., Udry, S., \& Pepe, F.\ 2015, \aap, 577, A62 
\bibitem[Yan \& Henning(2018)]{Yan18} Yan, F., \& Henning, T.\ 2018, Nature Astronomy, 2, 714 
\bibitem[Zechmeister et al.(2014)]{Zechmeister2014} Zechmeister, M., Anglada-Escud{\'e}, G., \& Reiners, A.\ 2014, \aap, 561, A59 
\end{thebibliography}
\end{document}